\documentclass[aps,prx,reprint,superscriptaddress]{revtex4-2}
\pdfoutput=1
\usepackage{easy-todo}
\usepackage[dvipsnames]{xcolor}
\usepackage{graphicx}
\usepackage{amsmath}
\usepackage{amsfonts}
\usepackage{latexsym}
\usepackage{amssymb}
\usepackage{microtype}
\usepackage{verbatim}
\usepackage{bbold}
\usepackage{hyperref}
\usepackage{cleveref}
\allowdisplaybreaks
\usepackage{multirow}
\usepackage{bbm}
\usepackage{braket}
\hypersetup{
  colorlinks   = true, 
  urlcolor     = blue, 
  linkcolor    = blue, 
  citecolor   = blue 
}

\definecolor{royalazure}{HTML}{033AA8}

\newcommand{\figeq}[2][1cm]{%
  \vcenter{\hbox{\includegraphics[width=#1]{#2}}}%
}
\newcommand{\pairs}{\,\figeq[8.118pt]{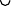} \dots \figeq[8.118pt]{Figures/cup}}

\newcommand{\smallsquare}{\mathord{\scalebox{0.7}{$\square$}}} 
\newcommand{\smallblacksquare}{\mathord{\scalebox{0.7}{$\blacksquare$}}}
\newcommand{\medcirc}{\mathord{\raisebox{-1pt}{\scalebox{1.4}{$\circ$}}}}
\newcommand{\medblackcirc}{\mathord{\raisebox{-1pt}{\scalebox{1.4}{$\bullet$}}}}
\newcommand{\sq}{\mathord{\scalebox{0.5}{$\square$}}} 

\begin{document}

\title{Free Cumulants and Full Eigenstate Thermalization from Boundary Scrambling}

\author{Felix Fritzsch}
\thanks{These authors contributed equally to this work.}
\affiliation{Max Planck Institute for the Physics of Complex Systems, 01187 Dresden, Germany}

\author{Gabriel O. Alves}
\thanks{These authors contributed equally to this work.}
\affiliation{Max Planck Institute for the Physics of Complex Systems, 01187 Dresden, Germany}

\author{Michael A. Rampp}
\affiliation{Max Planck Institute for the Physics of Complex Systems, 01187 Dresden, Germany}

\author{Pieter W. Claeys}
\affiliation{Max Planck Institute for the Physics of Complex Systems, 01187 Dresden, Germany}

\begin{abstract}
Out-of-time-order correlation functions (OTOCs) and their higher-order generalizations present important probes of quantum information dynamics and scrambling. 
We introduce a solvable many-body quantum model, which we term \textit{boundary scrambling}, for which the full dynamics of higher-order OTOCs is analytically tractable. 
These dynamics support a decomposition into free cumulants and unify recent extensions of the eigenstate thermalization hypothesis with predictions from random quantum circuit models. 
We obtain exact expressions for (higher-order) correlations between matrix elements and show these to be stable away from the solvable point.
The solvability is enabled by the identification of a higher-order Markovian influence matrix, capturing the effect of the full system on a local subsystem. 
These results provide insight into the emergence of random-matrix behavior from structured Floquet dynamics and show how techniques from free probability can be applied in the construction of exactly-solvable many-body models.
\end{abstract}

\maketitle

\section{Introduction} The relaxation of physical observables towards thermal equilibrium is a defining characteristic of the ergodic phase of matter and underlies the emergence of statistical mechanics in isolated quantum systems. 
The Eigenstate Thermalization Hypothesis (ETH) provides the leading theoretical framework for these phenomena, explaining how local subsystems thermalize by interpreting the remainder of the system as an effective bath~\cite{srednickiApproachThermalEquilibrium1999,deutschEigenstateThermalizationHypothesis2018,dalessio_quantum_2016}.
Recent progress has shifted the focus towards more refined probes of thermalization and quantum information dynamics. 
As a prominent example, out-of-time-order correlators (OTOCs) quantify the scrambling of quantum information~\cite{shenker_black_2014,maldacena2016bound,swingle_measuring_2016,swingle_unscrambling_2018,rakovszky_diffusive_2018,von_keyserlingk_operator_2018,nahum_operator_2018,khemani2018operator} and have been experimentally probed in analog~\cite{garttner_measuring_2017,li_measuring_2017,meier_exploring_2019,wei_emergent_2019} and digital quantum simulation~\cite{landsman_verified_2019,blok_quantum_2021,mi_information_2021,braumuller_probing_2022}. 
A higher-order generalization of the OTOC was recently measured on Google's 103-qubit quantum processor~\cite{abanin_constructive_2025}, where the complexity of a classical simulation of these dynamics motivated this experiment as a path to quantum advantage. 
In parallel with these developments, ETH has been extended to the so-called `full' ETH ansatz~\cite{foiniEigenstateThermalizationHypothesis2019a,brenesOutoftimeorderCorrelationsFine2021c,pappalardiEigenstateThermalizationHypothesis2022,jindalGeneralizedFreeCumulants2024,valliniLongtimeFreenessKicked2024,pappalardi_full_2025,fava_designs_2025,alves_probes_2025,fritzschMicrocanonicalFreeCumulants2024}, describing the dynamics of arbitrary multi-point correlators including (higher-order) OTOCs. 

This theory decomposes multi-point correlators in \emph{free cumulants}, connected correlation functions from free probability~\cite{voiculescuFreeNoncommutativeRandom1991,speicherFreeProbabilityTheory2003d,mingoFreeProbabilityRandom2017,nica2006lectures,novakThreeLecturesFree2012,speicherFreeProbabilityTheory2017b}, which recently found applications throughout many-body quantum dynamics~\cite{pappalardiEigenstateThermalizationHypothesis2022,hruza2023coherent,chen_free_2025,fava_designs_2025,jahnke2025free,camargo2025quantum,fritzsch_free_2025,dowling_free_2025}. 
Within full ETH, these free cumulants quantify correlations between eigenstates~\cite{pappalardiEigenstateThermalizationHypothesis2022}, which however lack a direct dynamical interpretation. 
Free cumulants were recently shown to emerge in a purely dynamical fashion in a minimal random circuit model, foregoing the notion of eigenstates entirely~\cite{fritzsch_free_2025}. 
From this perspective free cumulants effectively encode how different replicas of the time-evolution operator interact. 
These results were enabled by the identification of a higher-order influence matrix (IM).
Influence matrices capture the effect of a full system on a local subsystem, formalizing the notion of a bath~\cite{banuls2009matrix,lerose_influence_2021,sonner_influence_2021}, with a Markovian IM returning free cumulants as eigenmodes of the OTOC dynamics~\cite{fritzsch_free_2025}.

To bridge these two notions of free cumulants we introduce a new solvable model of many-body quantum dynamics, dubbed \emph{boundary scrambling}.
Beyond its intrinsic interest as an experimentally realizable model for which the dynamics of higher-order OTOCs is analytically tractable, this model 
(i) gives rise to Floquet dynamics and hence allows for a (quasienergy) eigenstate-based description and (ii) reproduces the generalized IM of the minimal random circuit model in the thermodynamic limit. 
Taken together, these allow for a unified analytical characterization of the free cumulants, with predictions for both eigenstate correlations and dynamical eigenmodes. 

\begin{figure}[ht!]
\includegraphics[width=0.95\columnwidth]{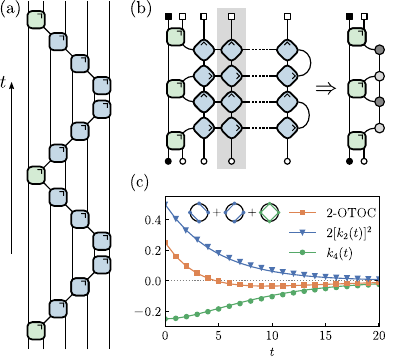}
\caption{Tensor network representation of (a) the boundary scrambling circuit and (b) a higher-point $k$-OTOC in the folded picture (left) and via the influence matrix (right). (c) Dynamics of the 2-OTOC and its constituting free cumulants, showing excellent agreement between finite-size numerics (markers) and analytical results in the thermodynamic limit (lines).} 
\label{fig:intro}
\end{figure}

The model describes spatially structured (1+1)D dynamics, with unitary evolution corresponding to a local quantum circuit model with a ``zig-zag" layout as shown in Fig.~\ref{fig:intro}(a). 
Higher-order OTOCs between local observables acting on the leftmost boundary site can be conveniently represented in the folded picture by the tensor network depicted in Fig.~\ref{fig:intro}(b).
The gates in the bulk can be compressed: Switching from time to space evolution and propagating the right boundary by means of the spatial transfer matrix (gray) yields a higher-order IM for the left boundary.
The solvability is underpinned by choosing the (blue) gates in the bulk to be dual-unitary, i.e., unitary in both time and space~\cite{bertini_exact_2019,gopalakrishnan_unitary_2019,bertini_exactly_2025}. 
The (green) gates at the boundary can be chosen freely.
In the thermodynamic limit the IM of this circuit coincides with the Markovian IM of the solvable random circuit model of Ref.~\cite{fritzsch_free_2025}.
The boundary scrambling circuit at the solvable point hence reproduces the solvable random-circuit dynamics of Ref.~\cite{fritzsch_free_2025} while circumventing the need for randomness and averaging over circuit realizations. 
This connection allows for an exact characterization of higher-order OTOCs and their decomposition in free cumulants, which now additionally encode correlations between eigenstates.
We emphasize that, in contrast to other analytically tractable settings, in addition to not requiring randomness the solvability requires neither a large number of local degrees of freedom nor late times ~\cite{roberts2015diagnosing,das2024late,maldacena2016remarks,von_keyserlingk_operator_2018,nahum_operator_2018,chan2018solution,khemani2018operator,yoshimura_operator_2025,chen_free_2025,bertini2020scrambling,claeys_maximum_2020}, which is what allows for making the connection with full ETH. 

This dynamics predicts that all higher-order OTOCs decay with the same rate, twice as fast as the time-ordered two-point correlation function, with all time scales analytically characterized.
We confirm this prediction with finite-size numerics and show that the eigenstate-based and dynamical free cumulants coincide. 
We obtain exact predictions for (higher-order) correlations between matrix elements.
For concreteness we focus on the standard 2-OTOC, for which we present explicit expressions for time- and frequency-resolved free cumulants, but we emphasize that our results extend to \emph{all} higher-order OTOCs. 
Remarkably, all OTOC dynamics is shown to be stable away from the solvable point.
While the proof of these results is relatively straightforward for the 2-OTOC, the extension to arbitrary-order $k$-OTOCs  ---as required for establishing free independence--- is significantly more involved and goes beyond standard applications of dual-unitarity, requiring the combination of tensor network manipulations with techniques from free probability.

\section{Model} 
\label{sec:model}
We consider a one-dimensional local quantum circuit with $L+1$ sites, each hosting a $d$-level qudit.
The left boundary (site 0) is the subsystem of interest, with the $L$ rightmost sites acting as an effective bath. 
The dynamics of the bath is governed by a ``zig-zag" circuit $\mathcal{V} = V_{1,2} V_{2,3} \cdots V_{L-1,L}V_{L-1,L}\cdots V_{2,3}V_{1,2}$ with $V_{ij}$ a dual-unitary two-qudit gate acting on sites $i$ and $j$.
The subsystem is coupled to the bath through a two-qudit gate $U = U_{0,1}$ which also induces the internal dynamics of the subsystem. 
The total evolution operator over a single discrete time step $\mathcal{U} = \left(\mathbf{1}_{0} \otimes \mathcal{V}\right)\left(U \otimes \mathbf{1}_{L-1}\right)$ gives rise to the Floquet dynamics of the model.

We focus on generalized OTOCs of arbitrary order $k$ between local observables $A$ and $B$ of the form $A=a \otimes \mathbf{1}_{L}$ and $B=b \otimes \mathbf{1}_{L}$, 
\begin{align}
\label{eq:OTOC_definition}
    C_k(t) = \frac{1}{D}\mathrm{tr}\left(\left[A(t)B\right]^k\right)  = \frac{1}{D} \mathrm{tr}\left(\left[\mathcal{U}^t A \,\mathcal{U}^{\dagger t} B\right]^k\right),
\end{align}
with $D=d^{L+1}$.
Assuming traceless $a$ and $b$ in the remainder of this work, a vanishing $k$-OTOC $C_k(t)$ after first taking the thermodynamic limit $L \to \infty$ and then sending $t\to \infty$ signals higher-order mixing and gives rise to \emph{free independence} between $A(t)$ and $B$~\cite{fava_designs_2025}.
The latter indicates maximum scrambling and, roughly speaking, statistical independence in the sense of a complete lack of algebraic relations between them. 

\section{Free cumulants} Full ETH predicts a decomposition of the $k$-OTOC $C_k(t)$ as a sum of terms indexed by noncrossing permutations, with each term factorizing in free cumulants corresponding to the size of the cycles in this permutations~\cite{foiniEigenstateThermalizationHypothesis2019a,pappalardiEigenstateThermalizationHypothesis2022,pappalardi_full_2025}. 
At lowest orders and for traceless observables, this decomposition returns
\begin{align}
\label{eq:moment_cumulant_formula}
C_1(t) = k_2(t) \quad \text{and} \quad C_2(t) = k_4(t) + 2k_2(t)^2 \,,
\end{align}
as also illustrated in Fig.~\ref{fig:intro}(c). 
These decompositions allow the free cumulants to be determined recusively, leading in particular to $k_4(t)= C_2(t) - 2C_1(t)^2$.
Full ETH predicts that these free cumulants can be expressed in terms of correlations between matrix elements $A_{ij}=\langle i|A|j\rangle$ of observables in the quasienergy eigenstates satisfying $\mathcal{U}\ket{i}=e^{\mathrm{i}\varphi_i}\ket{i}$.
The relevant lowest-order free cumulants are explicitly given by 
\begin{align}
    \label{eq:eth_cumulants}
    k_2(t) & = \frac{1}{D}\sum_{i\neq j}A_{ij}B_{ji}\,e^{\mathrm{i}\omega_{ij}t} \, , \nonumber \\
    k_4(t) & = \frac{1}{D}\sum_{i\neq j \neq k \neq l}A_{ij}B_{jk}A_{kl}B_{ji}\,e^{\mathrm{i}(\omega_{ij} + \omega_{kl})t}\,,
\end{align}
with $\omega_{ij} = \varphi_i - \varphi_j$ and the summation running over pairwise distinct indices only.
These free cumulants can also be expressed in the frequency domain via Fourier transform with respect to $t$, which effectively amounts to replacing $e^{\mathrm{i}\omega_{ij}t}$ by $2\pi\delta(\omega - \omega_{ij})$. 
In the frequency domain the free cumulants hence encode the dependence of correlations between matrix elements on the respective frequency $\omega$, i.e., the quasienergy difference.
For $A=B$ those correlations include the usual off-diagonal ETH in terms of $k_2(\omega)$, returning the spectral function quantifying the variance of matrix elements~\cite{dalessio_quantum_2016}, but extend the ETH paradigm when generalized to higher-order cumulants.
Note that $k_4(t)$ would vanish identically under the standard ETH assumption that off-diagonal matrix elements behave as i.i.d. Gaussian variables, such that this (nonvanishing) free cumulant captures correlations beyond ETH.

\subsection{Evolution in Space: Influence Matrix}
To contrast the eigenstate-based full ETH picture with the dynamical recursive description of free cumulants, we now obtain an exact description of the 2-OTOC dynamics, while referring to App.~\ref{app:proof} for the general case.
Utilizing a folding procedure we move into a replica picture for the OTOC dynamics, in which the OTOC is expressed in terms of folded gates
\begin{align}
    (U \otimes U^*)^{\otimes 2}=\figeq[0.065\columnwidth]{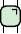} \, \, , \quad
    (V \otimes V^*)^{\otimes 2}=\figeq[0.12\columnwidth]{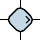} \, \, ,
\end{align}
with $*$ denoting complex conjugation. 
The appropriate boundary conditions are given by the permutation states $|\medcirc)=\figeq[0.038\columnwidth]{Figures/diag_id_state} = \sum_{i,j=1}^d \ket{iijj}$ and  $|\smallsquare)=\figeq[0.038\columnwidth]{Figures/diag_swap_state} = \sum_{i,j=1}^d \ket{ijji}$ corresponding to the identity and the swap permutation between the two replicas, respectively.
The observables are encoded by dressing these permutation states as $\figeq[0.038\columnwidth, angle=180]{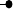} = |\medblackcirc) = (a \otimes \mathbf{1}_0 \otimes a \otimes \mathbf{1}_0)|\medcirc)$ and $\figeq[0.038\columnwidth, angle=180]{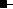} = (\smallblacksquare| =  (\smallsquare|(b \otimes \mathbf{1}_0 \otimes b \otimes \mathbf{1}_0)$.
Dual-unitarity results in the following graphical identities for the folded $V$~\cite{bertini_exactly_2025}:
\begin{align}\label{eq:dualunitarity_V}
\figeq[0.13\columnwidth]{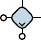} = 
\figeq[0.13\columnwidth]{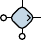} = 
\figeq[0.07\columnwidth]{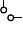} ,\,\,
\figeq[0.13\columnwidth]{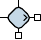} = 
\figeq[0.13\columnwidth]{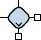} = 
\figeq[0.07\columnwidth]{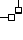} 
\end{align}

In the replica picture the OTOC is expressed as the compact tensor network shown in Fig.~\ref{fig:intro}(b), resembling a two-dimensional partition function.
In order to contract this tensor network, we use the well-established \emph{space-time duality} and consider the dynamics in space. 
We identify a spatial transfer matrix (rotated by $90^\circ$ for convenience)
\begin{align}
    \mathcal{T} = \frac{1}{d}\, \figeq[0.5\columnwidth]{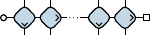}
\end{align}
and left and right spatial boundaries
\begin{align}
    &(\mathcal{B}|  = \frac{1}{d^2}\, \figeq[0.52\columnwidth]{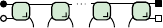} \\
    &|\pairs)  = \,\,\,\,\,\,\,\,\figeq[0.31\columnwidth]{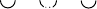} 
\end{align}
These are states defined on the temporal lattice of $2(t-1)$ sites, with each site carrying a $d^4$-dimensional (folded) Hilbert space. 
The full circuit can be recast as
\begin{align}\label{eq:C2_BTR}
    C_2(t) = ( \mathcal{B} | \mathcal{T}^{L-1} |\pairs) \equiv (\mathcal{B}|\mathcal{I}) \, .
\end{align}
All information about the local subsystem is contained in the left boundary $(\mathcal{B}|$, and $|\mathcal{I})=\mathcal{T}^{L-1} |\pairs)$ encodes the effect of the (replicated) bulk on local observables at this boundary.
The state $|\mathcal{I})$, dubbed the (generalized) IM, generalizes the IM approach to quantum dynamics established for time-ordered correlation functions~\cite{banuls2009matrix,lerose_influence_2021,sonner_influence_2021} to arbitrary-order OTOCs~\cite{fritzsch_free_2025}.

In the thermodynamic limit $L \to \infty$ this IM can be exactly obtained. 
In this limit $\mathcal{T}$ in Eq.~\eqref{eq:C2_BTR} can be replaced by a projector $\mathcal{P}$ onto its leading eigenspace, yielding $|\mathcal{I}) = \mathcal{P}|\pairs)$. 
As $\mathcal{T}$ is non-expanding, the leading eigenspace corresponds to eigenoperators with eigenvalue 1, and an exhaustive set of eigenvectors can be obtained using Eq.~\eqref{eq:dualunitarity_V} (App.~\ref{app:2OTOC}).
For the 2-OTOC the eigenvalue 1 is $(2t-1)$-fold degenerate, with eigenstates corresponding to domain wall configurations 
\begin{align}
|2(t-1),j) = \underbrace{\figeq[0.0225\columnwidth]{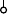} \cdots \figeq[0.0225\columnwidth]{Figures/circ_v}}_j \,\underbrace{ \figeq[0.0225\columnwidth]{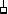} \cdots \figeq[0.0225\columnwidth]{Figures/sq_v}}_{2(t-1)-j}
\end{align}
switching between $\medcirc$ at (half) time step $j=0,\ldots,2(t-1)$ and $\smallsquare$ at $j+1$ (the so-called maximally chaotic subspace, see also Refs.~\cite{bertini_operator_i_2020,claeys_maximum_2020,bertini2020scrambling,rampp_dual_2023,huang_out--time-order_2023}).
Performing the appropriate orthonormalization and projection, the influence matrix follows as 
\begin{align}
    |\mathcal{I}) = &\frac{1}{d^{2(t-1)}}\!\sum_{j \text{ even}}|2(t-1),j)- \frac{1}{d^{2t-1}}\!\sum_{j \text{ odd}}|2(t-1),j),
    \label{eq:IM_2OTOC}
\end{align}
with weights depending on whether the domain wall is located at even or odd half time steps. Note that such domain wall configurations appear throughout the study of 2-OTOCs~\cite{bertini_operator_i_2020,claeys_maximum_2020,bertini2020scrambling,rampp_dual_2023,huang_out--time-order_2023,rakovszky_diffusive_2018,von_keyserlingk_operator_2018,nahum_operator_2018,Zhou2020,fisher_random_2023,jonay_physical_2024,yoshimura_operator_2025,suchsland2025dynamical}. 
As indicated in Fig.~\ref{fig:intro}(b) this IM can be cast in matrix-product form with a two-dimensional auxiliary space labelled by the permutations $\medcirc$ and $\smallsquare$, i.e.
\begin{align}
|\mathcal{I}) = \figeq[0.35\columnwidth]{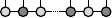}\,\,,
\end{align}
where the local tensors encode the locations and the weights of the domain walls, with non-zero components
\begin{align}
    \figeq[0.06\columnwidth]{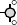} = 
    d^2 \figeq[0.06\columnwidth]{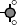} =
    \figeq[0.04\columnwidth]{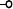}\, , \, 
    \figeq[0.06\columnwidth]{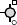} = 
    \figeq[0.06\columnwidth]{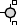} = 
    -d^3 \figeq[0.06\columnwidth]{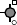} = 
    d^2 \figeq[0.06\columnwidth]{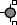} = 
    \figeq[0.04\columnwidth]{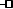}\, . 
\end{align}
The initial (final) boundary condition is obtained by summing the lower (upper) auxiliary leg over $\medcirc$ and $\smallsquare$. This influence matrix corresponds exactly to the influence matrix obtained for the 2-OTOC in Ref.~\cite{fritzsch_free_2025}. The 2-OTOC can be graphically represented as
\begin{align}\label{eq:OTOC_with_IM}
C_2(t) = \frac{1}{d^2}\,\,\figeq[0.52\columnwidth]{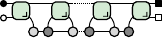}\, .
\end{align}

\subsection{Evolution in Time: Dynamical Semigroup} Having contracted the bulk into an influence matrix, we can now identify a temporal transfer matrix $\mathcal{G}$ acting as the generator of the dynamical semigroup.
From Eq.~\eqref{eq:OTOC_with_IM}, this transfer matrix reads
\begin{align}
    \mathcal{G} =  \figeq[0.125\columnwidth]{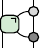} = \left( \begin{matrix}
       \mathcal{M}_{\sq\sq} & \quad \mathcal{M}_{\sq\circ} - \mathcal{M}_{\sq\sq}/d \\
       0 & \mathcal{M}_{\circ\circ}
    \end{matrix} \right),
\end{align}
here represented as block matrix in the auxiliary space $\{\smallsquare,\medcirc\}$.
The individual blocks are governed by operators $\mathcal{M}_{\sigma\nu}$ acting on (vectorized) $2$-replica observables
\begin{align}
\mathcal{M}_{\sq\sq} = \frac{1}{d^2}\figeq[0.085\columnwidth]{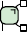}\,, \,\, \mathcal{M}_{\circ\circ} = \frac{1}{d^2}\figeq[0.085\columnwidth]{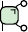}\,, \,\, \mathcal{M}_{\sq\circ} =\frac{1}{d^2}\figeq[0.085\columnwidth]{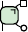}\,.
\end{align}
After including the appropriate boundary conditions, we obtain an expression for the 2-OTOC as
\begin{align}
    C_2(t) = (\psi_b|\mathcal{G}^t|\psi_a),
    \label{eq:2OTOC_semi_group}
\end{align}
with
\begin{align}
    (\psi_b| = \left(
    (\smallblacksquare|/d^2
    , 0 \right) \, , \quad
    |\psi_a) = \left( \begin{matrix}
       d \,\, |\medblackcirc) \\ 
       d^2 |\medblackcirc)
    \end{matrix} \right).
\end{align}
The matrix $\mathcal{G}$ acts on a $2 d^4$-dimensional Hilbert space, allowing for a straightforward numerical evaluation of the 2-OTOC for generic operators and arbitrary times $t$.

Analytical expressions can be obtained by noting that the diagonal channels $\mathcal{M}_{\sq\sq}$ and $\mathcal{M}_{\circ\circ}$ factorize in two copies of a quantum channel $\mathcal{M}$~\cite{fritzsch_free_2025}, defined as
\begin{align}
\label{eq:single_replica_channel}
\mathcal{M}(a) = \frac{1}{d}\mathrm{tr}_2 \left[U_{12} (a \otimes {\bf 1}) U_{12}^{\dagger}\right].
\end{align}
This channel determines the correlation functions as $C_1(t) = \mathrm{tr}(b \mathcal{M}^t(a))$~\cite{fritzsch_free_2025}. 
Choosing $a$ ($b$) as a left (right) eigenoperator of this channel with eigenvalue $\lambda$, we have that $C_1(t) = \lambda^t\, \mathrm{tr}(ab)/d$.
Furthermore, using $(\smallblacksquare|\mathcal{M}_{\sq\sq} = \lambda^2 (\smallblacksquare| $ and $\mathcal{M}_{\circ\circ} |\medblackcirc) = \lambda^2 |\medblackcirc)$, the 2-OTOC can be evaluated as
\begin{align}
\label{eq:dynamical_2OTOC}
 C_2(t) = \lambda^{2t-2} t\, (\smallblacksquare|\mathcal{M}_{\sq \circ}|\medblackcirc) -\lambda^{2t} (t-1)\frac{(\smallblacksquare|\medblackcirc)}{d} .
\end{align}
This expression can also be obtained by projecting these boundary conditions on the leading eigenoperators of $\mathcal{G}$ (see App.~\ref{app:2OTOC}), where the appearance of non-trivial Jordan blocks returns the polynomial prefactor $t$. This result indicates that the dynamics of Eq.~\eqref{eq:dynamical_2OTOC} captures the late-time dynamics of generic observables, albeit with modified prefactors due to the modified overlaps with eigenoperators.
We generally observe that the 2-OTOC asymptotically decays twice as fast as the two-point function, where the non-trivial Jordan structure of $\mathcal{G}$ gives rise to a linear correction to the overall exponential decay. 

The dynamical free cumulants follow as
\begin{align}\label{eq:dynamical_cumulants}
    k_2(t) &= k_2(a,b) \lambda^t \nonumber\\
    k_4(t) &= k_4(a,b,a,b) \lambda^{2 t} \nonumber\\
    &\qquad+t \lambda^{2(t-1)}
    \left[
        (\smallblacksquare|\mathcal{M}_{\sq \circ}|\medblackcirc) -
        \lambda^2(\smallblacksquare|\medblackcirc)/d
    \right],
\end{align}
with $k_2(a,b) = \mathrm{tr}(ab)/d $ and $k_4(a,b,a,b)=\mathrm{tr}(abab)/d - 2 (\mathrm{tr}(ab)/d)^2$ the (static) free cumulants from free probability~\footnote{Note that for $d=2$ and hermitian and traceless $a,b$ normalized as $\mathrm{tr}(a^2)/d = \mathrm{tr}(b^2)/d = 1$, $k_4(a,b,a,b) = -1/4$ does not depend on either $a$ or $b$. At time $t=0$ we hence find a universal value $k_4(t=0) = -1/4$.}. At short times the dynamics is governed by the exponential decay of $k_2(t)$, whereas at long times the OTOC is dominated by $k_4(t)$.

\begin{figure}
    \centering
    \includegraphics[width=\columnwidth]{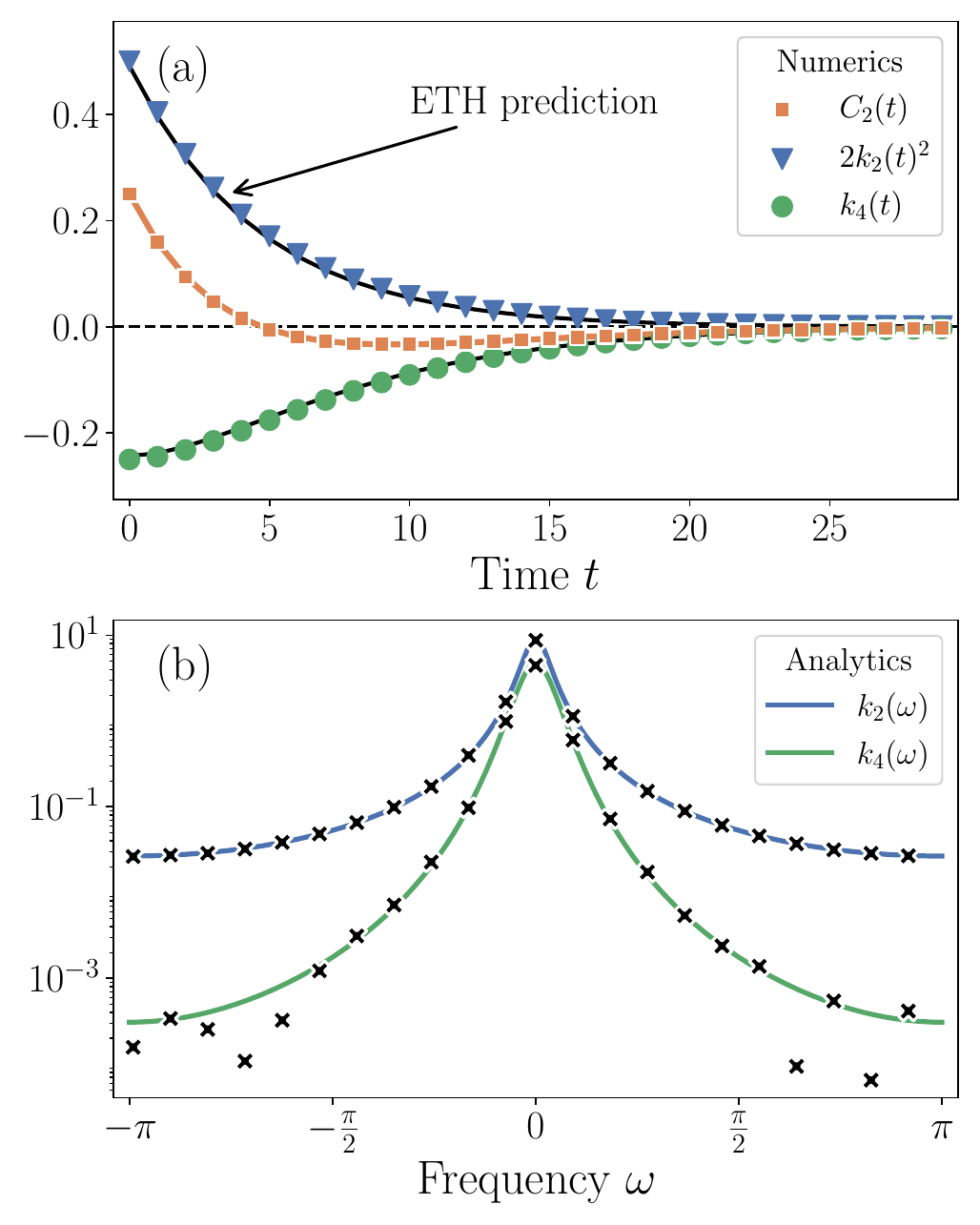}
    \caption{
    (a) Dynamics of the 2-OTOC and its decomposition in free cumulants $C_2(t) = k_4(t) + 2 [k_2(t)]^2$. 
	Analytic expressions [Eqs.~\eqref{eq:dynamical_2OTOC} and \eqref{eq:dynamical_cumulants}] (colored solid lines) are compared with numerical data for the free cumulants [Eq.~\eqref{eq:moment_cumulant_formula}] (markers) and the full-ETH predictions [Eq.~\eqref{eq:eth_cumulants}] obtained via exact numerical diagonalization (black dashed lines).
    (b) Frequency-resolved free cumulants. Numerical results from full ETH [Eq.~\eqref{eq:eth_cumulants}] (markers) are compared with analytical results [Eq.~\eqref{eq:frequency_cumulants}] (dashed lines).
    Numerical calculations performed for bath size $L = 10$ and $U$ chosen such that $\lambda \approx 0.9$.
    We plot the absolute value of the cumulants in log-scale to better visualize the long tails.
    }
    \label{fig:otoc_plot}
\end{figure}

\section{Numerics}
While the above exact results on the OTOC are obtained in the thermodynamic limit, they accurately capture the dynamics in finite systems even for moderate choices of $L$.
We consider a fixed choice of the unitary $U$, chosen to minimize finite-size effects, and random (in space, i.e., quenched disorder) dual-unitary gates $V$ (see App.~\ref{app:numerics} for details). 
We additionally choose $U$ to be Hermitian, such that the left and right eigenoperators of the quantum channel coincide and the free cumulants [Eq.~\eqref{eq:eth_cumulants}] encode correlations between matrix elements of a single operator $(A=B)$.
While quenched disorder is not necessary and the above derivation also applies for a homogeneous bulk of ergodic dual-unitary gates, the presence of disorder reduces finite-size effects.
As shown in Fig.~\ref{fig:otoc_plot}(a) we find almost perfect agreement between the finite-size numerics and our exact results for the 2-OTOC and the free cumulants [Eq.~\eqref{eq:dynamical_cumulants}].
By performing exact diagonalization of the evolution operator $\mathcal{U}$ we obtain the free cumulants via the explicit full-ETH prediction [Eq.~\eqref{eq:eth_cumulants}] and again find almost perfect agreement, see Fig.~\ref{fig:otoc_plot}(a).

A Fourier transform of Eq.~\eqref{eq:dynamical_cumulants} returns 
\begin{align}\label{eq:frequency_cumulants}
    k_2(\omega) 
    &= 
    k_2(a, b) J^{(1)}_\gamma(\omega), \nonumber\\
    k_4(\omega) 
    &=
    k_4(a, b, a, b) J^{(1)}_{2\gamma}(\omega)
    \nonumber \\ 
    &\qquad + 
    J^{(2)}_{2\gamma}(\omega)
    \left[
        (\smallblacksquare|\mathcal{M}_{\sq \circ}|\medblackcirc)/\lambda^2 
        -
        (\smallblacksquare|\medblackcirc)/d
    \right],
\end{align}
where we have introduced the decay timescale $\gamma$ through $\gamma \equiv -\ln |\lambda|$ and defined
\begin{align}
    J^{(1)}_{\gamma}(\omega)
    &=
    \frac{\sinh \gamma}{\cosh \gamma - \cos \omega}\,, \nonumber \\
    J^{(2)}_{\gamma}(\omega)
    &=
    \frac{\cos(\omega)\cosh(\gamma) - 1}{\left[\cos(\omega) - \cosh(\gamma)\right]^2}\,.
\end{align}
For slowly decaying modes with $\lambda$ close to one and hence small $\gamma$, $k_2(\omega)$ approximates Lorentzian correlations between matrix elements, consistent with recent predictions for dual-unitary circuits~\cite{fritzsch_eigenstate_2021}.

These results again agree with the numerically obtained full-ETH prediction~\eqref{eq:eth_cumulants}, see Fig.~\ref{fig:otoc_plot}(b), quantifying frequency-resolved correlations between matrix elements. 
Small deviations from the analytic result appear at the tail ends of $k_4(\omega)$ due to finite-size effects.
Higher-order cumulants can assume negative values and also drop off as a Lorentzian at large frequencies, consistent with recent numerics~\cite{fritzschMicrocanonicalFreeCumulants2024,pappalardi_full_2025}.
These expressions present a first analytical prediction for higher-order correlations between matrix elements, capturing correlations beyond ETH, and are expected to be asymptotically correct for generic observables.

\section{Higher-order OTOCs and stability}
All results can be extended towards $k$-OTOCs with $k>2$ (see App.~\ref{app:proof}). Crucially, the influence matrix of the bulk returns the random-circuit influence matrix at all orders $k$, such that the results of Ref.~\cite{fritzsch_free_2025} can be directly applied, again circumventing the need for randomness and averaging. 
At late times these results establish the asymptotic freeness of $A(t)$ and $B$ and the approach to free independence through the decay of free cumulants.

For the $k$-OTOC, the tensor network of Fig.~\ref{fig:intro}(b) is now constructed out of gates consisting of $k$-folded replicas, with the permutation states at the boundary corresponding to the identity ($\medcirc$) and the cyclic ($\smallsquare$) permutation of $k$ elements. 
The relevant eigenstates of the spatial transfer matrix no longer correspond to domain wall states, but rather to multichains: ordered sequences of noncrossing permutations. 
The theory of free probability is built on noncrossing permutations, which here determine the eigenstates of the spatial transfer matrix.
Such eigenstates were previously obtained in Ref.~\cite{chen_free_2025}, where their more complicated structure however prevented a direct orthonormalization.
Using identities from free probability, it is possible to establish the IM as an MPS of the form of Eq.~\eqref{eq:OTOC_with_IM}.
The auxiliary space is labelled by noncrossing permutations of $k$ elements, and the total number of such noncrossing permutations returns a bond dimension equal to the $k$-th Catalan number. 
The relevant tensors now correspond to permutation states in a $d^{2k}$-dimensional Hilbert space, defined as
\begin{align}
    \figeq[0.06\columnwidth]{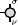} = 
    \figeq[0.06\columnwidth]{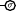}\,\, \delta_{\nu \subseteq \sigma}, \quad
    \figeq[0.06\columnwidth]{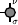} =  \figeq[0.06\columnwidth]{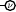}\,\, d^{-\ell(\sigma,\nu)} \mu(\sigma,\nu)\, \delta_{\sigma \subseteq \nu},  
\end{align}
The states and indices correspond to noncrossing permutations $\sigma,\nu \in \textrm{NC}(k)$, with $\sigma \subseteq \nu$ if all cycles of $\sigma$ are contained in the cycles of $\nu$. 
The permutation states are defined by a permutation $\sigma$ as $|\sigma) = \sum_{i_1, \dots i_k=1}^d |i_1, i_{\sigma(1)} \dots i_k,i_{\sigma(k)})$, generalizing the identity and swap permutation for $k=2$ to general noncrossing permutations.
The prefactor in the second tensor depends on $\ell(\sigma,\nu)$, i.e. the distance between $\sigma$ and $\nu$ defined as $\ell(\sigma,\nu) = k-|\sigma^{-1}\nu|$, with $|\sigma^{-1}\nu|$ the number of cycles in the permutation, and the M\"obius function of the noncrossing permutations $\mu(\sigma,\nu)$ (see App.~\ref{app:proof} for details).
The generator $\mathcal{G}$ of the dynamical semigroup can be similarly extended, and the asymptotic $k$-OTOC dynamics follows from the leading eigenmodes of this generator, argued in Ref.~\cite{fritzsch_free_2025} to correspond to free cumulants.

The repeated appearance of this higher-order influence matrix in both the structured boundary scrambling dynamics and the random circuit model suggests its more general applicability as a minimal Markovian bath for $k$-OTOCs.
However, as this influence matrix is obtained by projecting the boundary state into the (highly degenerate) leading eigenspace of the spatial transfer matrix $\mathcal{T}$, it is naively expected to be highly unstable away from the dual-unitary point. 
The number of multichains and hence the degeneracy of this eigenspace scales combinatorially with $t$ and $k$.
Away from dual-unitarity, this degeneracy is broken and numeric investigations show that the spatial transfer matrix has a unique leading eigenstate with eigenvalue one. 
From first-order perturbation theory we hence expect corrections arising from all states in this space.
This observation would be consistent with the general instability of solvable (e.g. integrable) dynamics away from solvable points.

Remarkably, the influence matrix is in fact perturbatively stable when moving away from dual-unitarity.
Similar in spirit to degenerate perturbation theory, the effect of small dual-unitarity-breaking perturbations can be taken into account by projecting the spatial transfer matrix $\mathcal{T}$ into the eigenspace at the dual-unitary point and diagonalizing this projected (non-dual-unitary) transfer matrix.
In App.~\ref{app:proof}, we analytically show that the spatial transfer matrix $\mathcal{T}$ for \emph{generic} unitary gates projected into the space of multichains retains the exact influence matrix $|\mathcal{I})$ as (generically unique) eigenstate with eigenvalue one:
\begin{align}
    \mathcal{P} \mathcal{T} \mathcal{P}|\mathcal{I}) =  \mathcal{P} \mathcal{T} |\mathcal{I}) = |\mathcal{I}),
\end{align}
where $\mathcal{P}$ is the projector on the multichain states, i.e. the eigenstates with eigenvalue one at the dual-unitary point. 
The influence matrix hence only receives corrections from non-multichain states, which are suppressed by the gap of $\mathcal{T}$.
For gates in the bulk close to dual-unitarity the $k$-OTOC hence satisfies $C_k(t) \approx ( \mathcal{B}|\mathcal{I})$, with corrections suppressed as we approach the dual-unitary point.
We illustrate this result with further numerics in App.~\ref{app:stability_numerics}. 
For small deviations from dual-unitarity and large system sizes, the OTOC dynamics is visually indistinguishable from the dual-unitary result.
While the structural stability of dual-unitary dynamics is in general an open question~\cite{kos_correlations_2021,ippoliti_dynamical_2023,rampp_dual_2023,riddell2024structural,fritzsch_free_2025}, this result establishes the stability of the $k$-OTOC dynamics in the present model.

\section{Discussion and Conclusions}
We have introduced the boundary scrambling circuit as a first example of a spatially structured, chaotic, yet exactly solvable many-body Floquet system in which the dynamics of higher-order OTOCs can be fully characterized.
Using this exact solution we can verify predictions from recent extensions of ETH, fully characterize free cumulants as the building block of higher-order correlators, and establish asymptotic freeness between static and dynamic observables. 
The exact results for the dynamics additionally allow us to obtain frequency-resolved expressions for higher-order correlations between matrix elements, exhibiting a characteristic Lorentzian form.
The presented results are expected to hold asymptotically in more generic ergodic quantum dynamics.
Next to these results on full ETH, this work presents the first structured model in which a Markovian influence matrix can be established on the level of $k$-OTOCs, following the success of the influence matrix approach in describing the dynamics of correlation functions of structured models~\cite{banuls2009matrix,lerose_influence_2021,sonner_influence_2021}.

The solvability of this model should be contrasted with the solvability of dual-unitary brickwork circuits (as recently reviewed in Ref.~\cite{bertini_exactly_2025}).
Dual-unitarity has served as a testbed for notions of quantum chaos, e.g. through spectral statistics and emergent designs, but results on (higher-order) OTOCs typically require a scaling limit and no exact results are available for their full dynamics. 
Boundary scrambling, conversely, allows for a complete characterization of $k$-OTOCs by effectively inducing a separation of time scales between the boundary dynamics and the bulk dynamics, where in each time step the bulk \emph{globally} scrambles quantum information.
It is expected that most of the exact results for dual-unitary brickwork circuits can be extended to the boundary scrambling geometry, where it would be interesting to contrast freeness with different probes of quantum chaos.
A similar question was raised in Ref.~\cite{ippoliti_infinite_2025}, which introduced a model based on linear feedback shift registers that provably satisfied ETH.
A recent study of random matrix product unitaries additionally indicated how freeness needs to be distinguished from the formation of unitary designs as an indicator of quantum chaos~\cite{dowling_free_2025}, and proposed freeness as an operational benchmark in quantum devices. 
Boundary scrambling here presents a structured model which reproduces the mathematical structure of random matrix product unitaries, and can also be directly implemented in current digital quantum simulation platforms.

Our results can be extended in different directions.
This Markovian influence matrix presents a first step towards characterizing non-Markovianity in higher-order quantum memory.
Non-Markovianity can be introduced in a controlled way in the boundary scrambling model by e.g. considering smaller system sizes or stronger deviations from dual-unitarity. 
Furthermore, the presented results can be extended to more structured subsystems: the influence matrix approach does not depend on any particular choice of subsystem.
Here it would also be interesting to study higher-order OTOCs for spatially separated observables, following Ref.~\cite{chen_free_2025}, where the application of tools from free probability can provide further insight into the structure of information spreading in extended systems.
The introduction of the influence matrix also effectively modifies the unitary circuit dynamics to non-unitary dynamics determined by the (non-Hermitian, non-unitary) generator of the dynamical semigroup. 
Such a move is reminiscent of recent research on Ruelle-Pollicott resonances~\cite{prosen_ruelle_2002,prosen_ruelle_2004,garcia-mata_chaos_2018,znidaric_momentum-dependent_2024,mori_liouvillian-gap_2024,zhang_thermalization_2025,jacoby_spectral_2025,duh_ruelle-pollicott_2025}, suggesting the influence matrix approach as a general probe of (higher-order) Ruelle-Pollicott resonances.
To conclude, we expect that similar techniques can be applied for obtaining the exact dynamics of other multi-replica observables, i.e., nonlinear functions of local observables or states as well as spectral fluctuations and eigenstate correlations. 

\begin{acknowledgements}
We acknowledge support from the Max Planck Society.
F.F. acknowledges support from the European Union's Horizon Europe program under the Marie Sk{\l}odowska Curie Action GETQuantum (Grant No. 101146632).
\end{acknowledgements}

\appendix

\bibliography{Library_FP}

\widetext

\section*{APPENDICES}

In Appendix~\ref{app:proof} we prove that the boundary scrambler returns the random-circuit influence matrix at arbitrary orders of the $k$-OTOC and that it is stable when perturbing away from dual-unitarity. 
Appendix~\ref{app:2OTOC} illustrates this proof with explicit examples and relevant derivations for the 2-OTOC.
Details of the numerics are provided in Appendix~\ref{app:numerics}.
In App.~\ref{app:stability_numerics} we provide further details on the stability away from dual-unitarity with some further numerical results.
Finally, in App.~\ref{app:concentration} we expand on how the system approach the thermodynamic limit and the analytical predictions with some extra numerics.

\section{Analytical derivation of the influence matrix}
\label{app:proof}
In this Appendix, we analytically show that the influence matrix from the main text presents the projection of the right boundary on the leading eigenspace of the transfer matrix when the bulk gates are dual-unitary. Away from the dual-unitary limit, we show that the transfer matrix projected on this leading eigenspace retains the influence matrix as leading eigenvector.

We first reintroduce the basic concepts and definitions in order to be self-contained. We then state the main theorem, with two corollaries returning the results mentioned above. In order to prove the main theorem, we first establish a lemma which combines tensor network manipulations with identities from free probability. This lemma allows for the local `propagation' of singletons, reminiscent of dual-unitary calculations -- but applies for generic choices of unitary gates.
We conclude by explicitly illustrating the main results in the specific case of the $2$-OTOC. \\

\textbf{Basic concepts.} 
We consider states $|\sigma)$ labelled by noncrossing permutations $\sigma \in \textrm{NC}(k)$, which act on $2k$ copies of a local $d$-dimensional Hilbert space as
\begin{align}
 |\sigma) =  \,\,\figeq[0.027\columnwidth]{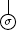}\qquad \textrm{with} \qquad  (i_1,i_1',i_2,i_2', \dots i_k,i_k'|\sigma) = \prod_{j=1}^k \delta_{i_j',i_{\sigma(j)}}\,, \quad i_j,i_j'=1\dots d.
\end{align}
These states corresponds to folded versions of permutation operators, permuting $k$ Hilbert spaces according to the permutation $\sigma$. The overlap between two such permutation states is given by
\begin{align}
  (\nu|\sigma) = \,\, \figeq[0.027\columnwidth]{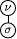}\,\, = d^{|\nu^{-1}\sigma|}\,,
\end{align}
with $|\nu^{-1}\sigma|$ the numbers of cycles in $\nu^{-1}\sigma$. This number of cycles defines a metric on the space of permutations, where the distance $\ell$ between $\nu,\sigma \in \textrm{NC}(k)$ is defined as $\ell(\nu,\sigma) = k - |\nu^{-1}\sigma|$.

The noncrossing permutations support an ordering, where $\sigma_i \subseteq \sigma_j$ if all cycles of $\sigma_i$ are contained in the cycles of $\sigma_j$. Note that all noncrossing permutations lie between the identity and the cyclic permutation, i.e. $\circ \subseteq \sigma \subseteq \smallsquare$. The distances satisfy a geodesic condition $\ell(\sigma_1,\sigma_2)+\ell(\sigma_2,\sigma_3) = \ell(\sigma_1,\sigma_3)$ if $\sigma_1 \subseteq \sigma_2 \subseteq \sigma_3$.
With this ordering, the set of noncrossing permutations forms a partially ordered set, with a M\"obius function $\mu(\sigma_i,\sigma_j)$ defined as satisfying
\begin{align}
    \sum_{\sigma_1 \subseteq \sigma_2 \subseteq \sigma_3 }\mu(\sigma_2,\sigma_3) = \delta_{\sigma_1,\sigma_3},
\end{align}
where the summation runs over $\sigma_2$ with fixed $\sigma_1$ and $\sigma_3$. Written out explicitly, the M\"obius function reads
\begin{align}\label{eq:app:mobius}
    \mu(\sigma,\nu) = \prod_{V \in \sigma^{-1}\nu} (-1)^{|V|-1}C_{|V|-1},
\end{align}
where the product runs over the cycles $V$ of $\sigma^{-1}\nu$, with length $|V|$, and with $C_m$ the Catalan numbers.
States indexed by so-called multichains of noncrossing permutations, i.e. a sequence of noncrossing permutations $\vec{\sigma} = \sigma_1 \subseteq \sigma_2 \subseteq \dots \subseteq \sigma_{2\tau}$, can be constructed as the direct product of $2 \tau$ permutation states,
\begin{align}
    |\vec{\sigma}) = |\sigma_1) \otimes |\sigma_2) \otimes \dots \otimes |\sigma_{2 \tau}) = \,\,\,\figeq[0.205\columnwidth]{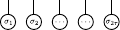}\,\,.
\end{align}
We will refer to multichains and the states they index interchangeably. The overlap between two multichains follows as
\begin{align}
    (\vec{\nu}|\vec{\sigma}) = \prod_{i = 1}^{2\tau} d^{|\nu_i^{-1} \sigma_i|}\,.
\end{align}
The spatial transfer matrix $\mathcal{T}$ is defined as
\begin{align}
    \mathcal{T} =  \frac{1}{d}\,\, \figeq[0.25\columnwidth]{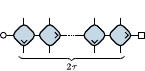}\,\,,
\end{align}
here rotated by $90^{\circ}$ for ease of notation. Each of the gates represents $k$ replicas of the unitary gate $U$ and its complex conjugate as $(U \otimes U^*)^{\otimes k}$. The right boundary state can similarly be represented as
\begin{align}
    |\pairs) = \figeq[0.185\columnwidth]{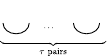}\,\,,
\end{align}
corresponding to a product of Bell pairs in the composite Hilbert space of $k$ replicas. The overlap of this state with the multichains can be obtained as
\begin{align}
    (\vec{\sigma}|\pairs) = \figeq[0.265\columnwidth]{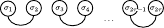} = \prod_{i=1}^{\tau} d^{|\sigma_{2i-1}^{-1}\sigma_{2i}|} = \prod_{i=1}^{\tau} d^{|\sigma_{2i}|-|\sigma_{2i-1}|+k}\,\,.
\end{align}

\textbf{Definition.}
The influence matrix for $t \equiv \tau+1$ times steps, which we repeat here for convenience, is defined as
\begin{align}\label{eq:app:IM}
  |\mathcal{I}) = \sum_{\vec{\sigma} } \left(\prod_{i=1}^{\tau}\mu(\sigma_{2i-1}, \sigma_{2i})\,d^{|\sigma_{2i}|-|\sigma_{2i-1}| -k }\right)|\vec \sigma).
\end{align}
where the summation runs over all multichains of $2\tau$ noncrossing permutations, $\vec{\sigma} =  \sigma_1 \subseteq \sigma_2 \subseteq \dots \subseteq \sigma_{2(t-1)}$ with $\sigma_i \in \textrm{NC}(k), \forall i$.
In the dual-unitary case, these multichains are left and right eigenstates of the spatial transfer matrix $\mathcal{T}$ with unit eigenvalue, $\mathcal{T}|\vec{\sigma}) = |\vec{\sigma})$ and $(\vec{\sigma}|\mathcal{T} = (\vec{\sigma}|$, as originally established in Ref.~\cite{chen_free_2025}\footnote{While Ref.~\cite{chen_free_2025} established these eigenstates for a transfer matrix in which the gates did not have the alternating orientation of the spatial transfer matrix of this work, this difference is immaterial for dual-unitary gates.}. \\

\textbf{Theorem.}
The main result in this Appendix is a proof for the following identity, where $\mathcal{T}$ is the spatial transfer matrix for \emph{any} choice of unitary gates and $|R)$ is the spatial boundary vector:
\begin{align}\label{eq:app:identity}
   (\vec{\nu}| \mathcal{T} |\mathcal{I}) = (\vec{\nu}|\pairs), \quad \forall \vec{\nu}. 
\end{align}
Writing out the matrix elements of the transfer matrix $\mathcal{T}$, Eq.~\eqref{eq:app:identity} reads
\begin{align}\label{eq:app:identity_matel}
  \frac{1}{d}\sum_{\vec{\sigma} } \left(\prod_{i=1}^{\tau}\mu(\sigma_{2i-1}, \sigma_{2i})\,d^{-k+|\sigma_{2i}|-|\sigma_{2i-1}|}\right) \figeq[0.25\columnwidth]{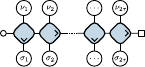} = \prod_{i=1}^{\tau} d^{|\nu_{2i}|-|\nu_{2i-1}|+k}.
\end{align}

{\bf Corollary 1.} The influence matrix $\mathcal{I}$ corresponds to $|\pairs)$ projected on the space of multichains $\{|\vec{\nu})\}$. Denoting this projector as $\mathcal{P}$, we find that
\begin{align}
    \mathcal{P}|\pairs) = |\mathcal{I}).
\end{align}

{\it Proof.} In the specific case where the constituting gates are dual-unitary, the multichains are right eigenstates of $\mathcal{T}$ and Eq.~\eqref{eq:app:identity} implies that $(\vec{\nu}| \mathcal{T} |\mathcal{I}) = (\vec{\nu}|\mathcal{I}) = (\vec{\nu}|\pairs)$. The projected state $\mathcal{P}|\pairs)$ is uniquely defined as a linear combination of multichains which satisfies $(\vec{\nu}|\mathcal{P}|\pairs) = (\vec{\nu}|\pairs)$. Both conditions are satisfied by $|\mathcal{I})$, the former by definition and the latter from the previous argument. Note that this final identity is independent of the choice of gates.\\

{\bf Corollary 2.} The influence matrix is an eigenvector of the transfer matrix for generic unitary gates when projected into the space of multichains. 

{\it Proof.} For generic choices of unitary gates, we can rewrite Eq.~\eqref{eq:app:identity} as
\begin{align}
       (\vec{\nu}| \mathcal{P} \mathcal{T}  \mathcal{P} |\mathcal{I}) = (\vec{\nu}|\mathcal{I}), \quad \forall \vec{\nu} \quad \Rightarrow \quad   \mathcal{P} \mathcal{T}  \mathcal{P} |\mathcal{I}) = \mathcal{P} \mathcal{T} |\mathcal{I}) = |\mathcal{I}),
\end{align}
directly returning the corollary since both sides have identical overlaps with all multichains and do not act outside of the space of multichains. \\

The proof of the theorem proceeds by grouping together different terms in the summation leading to the same matrix element $(\vec{\nu}|\mathcal{T}|\vec{\sigma})$, up to possible prefactors of $d$, and showing that they cancel. In the noncancelling terms unitarity can be used to repeatedly remove unitary gates and simplify this matrix element to a (product of) simple contraction(s), i.e. loops which each return a factor $d$, returning the right-hand side. In order to perform this grouping, we first establish the following lemma. \\

{\bf Lemma.} For a fixed noncrossing partition $\sigma_2$ and a fixed noncrossing permutation $\nu_1$ with a singleton $(n)$, i.e. $n$ forms a cycle of length 1 in $\nu_1$, the following identity holds:
\begin{align}\label{eq:app:lemma}
    \sum_{\sigma_1 \subseteq \sigma_2} \mu(\sigma_{1}, \sigma_{2})\,d^{|\sigma_2|-|\sigma_1|-k } \,\,\figeq[0.09\columnwidth]{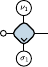} = 0 \qquad \textrm{unless $(n)$ is a singleton in $\sigma_2$.} 
\end{align}

{\it Proof.} We first note that if $\nu_1$ has a singleton $(n)$, which is necessarily also a singleton in the identity partition $\circ$,  then we can use unitarity in the $n$-th replica to contract a single unitary gate with its Hermitian conjugate. Consider e.g. $\nu_1 = (1)(23)$ and $n=1$, then unfolding the gate results in
\begin{align}
    \figeq[0.11\columnwidth]{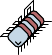} = \figeq[0.11\columnwidth]{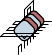} \,.
\end{align}
Note that here the blue gates no longer denote folded/replicated gates but represent a single gate $U$, whereas the red gates denote $U^*$.
Unitarity can be used to contract the two gates $U$ and $U^*$ connected by the singleton.
Second, we note that all permutations $\sigma_1$ which are identical after removing $n$ (`tracing out $n$') return the same contraction at the leftmost gate, up to a possible prefactor of $d$. We again illustrate this for $\nu_1 = (1)(23)$ and $n=1$, where $\sigma_1$ either $(12)(3)$ or $(1)(2)(3)$ returns the same contraction:
\begin{align}\label{eq:app:lemma_unfolding}
    \sigma_1 = (12)(3): \quad \figeq[0.11\columnwidth]{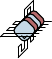} = \figeq[0.11\columnwidth]{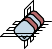}, \qquad \sigma_1 = (1)(2)(3): \quad  \figeq[0.11\columnwidth]{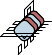} = d \figeq[0.12\columnwidth]{Figures/fig_app_unfolded_3}\,\,.
\end{align}
Note that if $n$ is a singleton in $\sigma_1$, we obtain a prefactor $d$ from the single loop, whereas if $n$ is not a singleton in $\sigma_1$ there is no additional prefactor.

We now separate the summation over $\sigma_1$ in Eq.~\eqref{eq:app:lemma} in equivalency classes based on this singleton structure, where we say that $\sigma_1 \stackrel{n}{\equiv} \sigma'_1$ if both are identical after tracing out $n$. For permutations within the same equivalence class, we observe that (i) the operator part of Eq.~\eqref{eq:app:lemma} is identical, and (ii) these have the same prefactor power of $d$. The former was established above. For the latter, we note that if $n$ is a singleton in $\sigma_1$ and $\sigma_1 \stackrel{n}{\equiv} \sigma'_1$, where $\sigma'_1 \neq \sigma_1$ such that $n$ is not a singleton in $\sigma'_1$, then $|\sigma'_1| = |\sigma_1|-1$. If $(n)$ is not a singleton in $\sigma_1'$ it is necessarily part of a larger cycle, e.g., if $n=1$ and $\sigma_1 = (1)(2)(34)$, then $\sigma_1'$ is either $(12)(34)$ or $(134)(2)$. 
The additional prefactor of $d$ from Eq.~\eqref{eq:app:lemma_unfolding} when $n$ is a singleton in $\sigma_1$ hence cancels with the additional factor of $d^{-1}$ from the additional cycle in $\sigma_1$ in the overall prefactor $d^{|\sigma_2| - |\sigma_1|-k}$, such that all permutations in the same equivalence class have the same prefactor.

To conclude the proof of this lemma, we establish the following crucial identity. Different terms in the summation in Eq.~\eqref{eq:app:lemma} can be grouped according to their equivalence class, and the summation over each class vanishes identically unless this equivalency class is trivial and consists of a single partition with $n$ as singleton. For fixed $\sigma_1$ and $\sigma_2$, the summation over all $\sigma_1'$ that are equivalent to $\sigma_1$ vanishes unless $n$ is a singleton in $\sigma_2$ and hence in all permutations in the summation. We have that, for fixed $\sigma_1$ and $\sigma_2$,
\begin{align}
    \sum_{\substack{\sigma_1' \subseteq \sigma_2 \\ \sigma_1' \stackrel{n}{\equiv} \sigma_1}} \mu(\sigma_{1}', \sigma_{2}) = 0 \qquad \textrm{unless $n$ is a singleton in $\sigma_2$}.
\end{align}
This identity can be proven by using the explicit parametrization of the M\"obius function. The M\"obius function [Eq.~\eqref{eq:app:mobius}] reads
\begin{align}
\mu(\sigma_1,\sigma_2) = \prod_{V \, \in \, \sigma_1^{-1}\sigma_2} (-1)^{|V|-1} C_{|V|-1}\,,
\end{align}
where $V$ denotes the cycles in $\sigma_1^{-1}\sigma_2$, with $|V|$ the length of the cycle, and $C_m$ are the Catalan numbers.
If $\sigma_2$ has a cycle of length $\ell$ containing $n$, then only the factor involving this cycle will differ across different $\sigma_1$'s within the same equivalency class. We take this cycle to be of length $\ell$ and relabel its elements as $(1 2 \dots \ell) = \smallsquare$. We define $\pi_1$ as $\sigma_1$ restricted to these $\ell$ elements, where since $\sigma_1 \subseteq \sigma_2$ this singles out the relevant cycles of $\sigma_1$ without `cutting' any cycles.

In the M\"obius function, the relevant factor corresponds to $\pi_1^{-1} \smallsquare = \pi_1^{\ast}$. This partition corresponds to the Kreweras complement of $\pi_1$, which is the largest noncrossing partition that can be constructed on the dual of the partition. Within the same equivalence class (now of noncrossing permutations of $\ell$ elements), we denote $\pi$ as the noncrossing partition that has a singleton $n$, and use $\rho$ to denote the equivalent permutations in which $n$ is not a singleton. The identity to be proven reduces to
\begin{align}\label{eq:app:lemma_sq}
    \mu(\pi,\smallsquare) + \sum_{\rho \stackrel{n}{\equiv} \pi} \mu(\rho, \smallsquare) = 0 \qquad \textrm{unless $n$ is a singleton in $\sigma_2$}.
\end{align}
The proof can be performed by graphically illustrating how the different $\rho$'s can be constructed from $\pi$ on the level of the Kreweras complement, which directly relates to the M\"obius function. We first illustrate this construction using a concrete example, choosing $\ell = 6$, $n=1$ and $\pi = (1)(23)(4)(56)$. The three corresponding choices of $\rho$ are then $\rho_1 = (123)(4)(56)$, $\rho_2 = (14)(23)(56)$ and $\rho_3 = (156)(23)(4)$. Illustrating these permutations and their Kreweras complement, we find that
\begin{align}
    &\figeq[0.11\columnwidth]{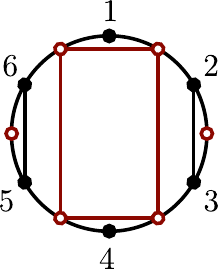}& \quad
    &\figeq[0.11\columnwidth]{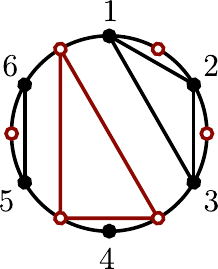}& \quad
    &\figeq[0.11\columnwidth]{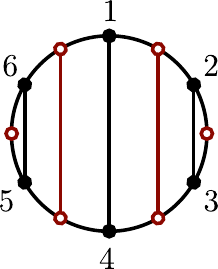}& \quad
    &\figeq[0.11\columnwidth]{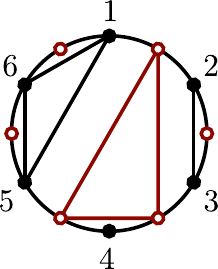}& \\
    & (1)(23)(4)(56)&  &(123)(4)(56)& &(14)(23)(56)& &(156)(23)(4)& \nonumber\\
    &\mu(\pi, \smallsquare) = -5 & &\mu(\rho_1,\smallsquare) = 2& &\mu(\rho_2,\smallsquare) =1& &\mu(\rho_3,\smallsquare) = 2& \nonumber
\end{align}
The M\"obius function factorizes according to the cycles in the Kreweras complement, illustrated in red, where  cycles of length $1,2,3,4, \dots$ contribute a factor  $1,-1,2,-5, \dots$
It can be observed that these M\"obius functions indeed sum to zero. This example illustrates the general idea that, if $\pi^*$ has a cycle of length $p$ which contains $(\dots n \, n+1 \dots)$, which is guaranteed if $\pi$ has a singleton $n$, the Kreweras complement of $\rho$ can be obtained by splitting this single cycle in two (noncrossing) cycles of length $r$ and length $p-r$. All other cycles in $\pi^*$ are identical to cycles in $\rho^*$, such that these can again be factored out in the summation over M\"obius functions. Explicitly writing out the relevant factor originating from this cycle in Eq.~\eqref{eq:app:lemma_sq} returns
\begin{align}
    (-1)^{p-1} C_{p-1} + \sum_{r=1}^{p-1} (-1)^{r-1} C_{r-1} \times (-1)^{p-r-1} C_{p-r-1} = (-1)^{p-1}\left[ C_{p-1} - \sum_{r=1}^{p-1}C_{r-1}C_{p-r-1} \right]= 0 \quad \textrm{if $p \neq 1$}, 
\end{align}
which vanishes by using the recursion relation of the Catalan numbers~\cite{stanley_catalan_2015}.
We arrive at the final result: the summation over M\"obius functions vanishes unless the singletons of $\nu_1$ are also singletons of $\sigma_2$ and hence of $\sigma_1$.  \\

This argument can now be extended to relate the singletons in $\nu_{2i-1}$ to the singletons in $\sigma_{2i}$, which also correspond to singletons in all underlying permutations $\sigma_1 \subseteq \sigma_2 \subseteq \dots \subseteq \sigma_{2i}$. Furthermore, rather than using this identity to `propagate' singletons in $\nu_{2i-1}$ from the left, a similar identity can be applied to `propagate' singletons in $\nu_{2i}^*$ from the right.\\

{\bf Corollary.} For a fixed noncrossing permutation $\nu_{2i-1}$ with a singleton $(n)$, i.e. $n$ forms a cycle of length 1 in $\nu_{2i-1}$, and a fixed noncrossing partition $\sigma_{2i}$, the following identity holds:
\begin{align}\label{eq:app:cont_left}
    \sum_{\vec{\sigma}} \left(\prod_{j=1}^i \mu(\sigma_{2j-1}, \sigma_{2j})\,d^{|\sigma_{2j}|-|\sigma_{2j-1}|-k}\right) \figeq[0.24\columnwidth]{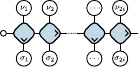}  = 0 \qquad \textrm{unless $(n)$ is a singleton in $\sigma_{2i}$,} 
\end{align}
provided $\nu_1 \subseteq \nu_2 \subseteq \dots \subseteq \nu_{2i-1}$ and where the summation again runs over multichains. Furthermore, this expression can be factorized: the $n$th replica acts as a singleton, whereas the remaining summation corresponds to Eq.~\eqref{eq:app:cont_left} for $k-1$ replicas (up to a possible prefactor of $d$). 

{\it Proof.} If $n$ is a singleton in $\nu_{2i-1}$ then the ordering of noncrossing permutations $\nu_1 \subseteq \nu_2 \subset \dots \subseteq \nu_{2i-1}$ implies that $n$ is also a singleton in these permutations. Furthermore, using the previous lemma, it follows that $\sigma_1$ and $\sigma_2$ share this singleton. Using this singleton structure, we can repeatedly use unitarity to contract the corresponding unitaries in the $n$th replica, which introduces a singleton along the `horizontal' direction. Such a singleton again introduces equivalence classes, and we can repeat the previous argument iteratively. The factorization directly follows from using unitarity in the $n$th replica, noting that removing the singleton $(n)$ from all cycles returns the corresponding expression for $k-1$ replicas, where the additional prefactor of $d$ can arise from the overlap between this singleton and $\nu_{2i}$. \\

{\bf Corollary.} For a fixed noncrossing permutation $\nu_{2i}$ with a singleton $(n)$ in its complement $\nu_{2i}^*$, i.e. $\nu_{2i}$ has a cycle containing $(\dots n\, n+1 \dots)$, and a fixed noncrossing partition $\sigma_{2i-1}$, the following identity holds:
\begin{align}
    \sum_{\vec{\sigma}} \left(\prod_{j=i}^{\tau} \mu(\sigma_{2j-1}, \sigma_{2j})\,d^{|\sigma_{2j}|-|\sigma_{2j-1}|-k}\right) \figeq[0.24\columnwidth]{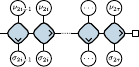} = 0 \qquad \textrm{unless $(n)$ is a singleton in $\sigma^*_{2i-1}$,} 
\end{align}
provided $\nu_{2i-1}\subseteq \nu_{2i} \subseteq \dots \subseteq \nu_{2\tau}$ and where the summation again runs over multichains. The resulting expression again factorizes, where the two replicas connected by this singleton factorize out.

{\it Proof.} This identity follows directly from the previous one by relabeling the different replicas in such a way that the cyclic permutation gets mapped to the identity permutation. This relabeling corresponds to taking the Kreweras complement, which reverses the ordering of the noncrossing permutations.\\

Given this lemma and the resulting corollaries, we are now in a position to prove Eq.~\eqref{eq:app:identity_matel} and hence the main theorem of this section. \\

{\it Proof of the theorem.} We use that any partition $\sigma$ either has a singleton or its complement $\sigma^*$ has a singleton. Using the lemma and its corollaries, and restricting the summation to be nonvanishing, this lemma can be applied at least once for every noncrossing partition $\nu_{i}$, either from the left or the right. Applying this lemma once effectively removes a single unitary and its hermitian conjugate from the replicas at each step, replacing them by a contraction. The resulting expression corresponds to a summation of the form of Eq.~\eqref{eq:app:identity_matel}, but now on $k-1$ replicas, with a possible prefactor depending on the Hilbert space dimension, and we can repeatedly use this argument to remove layers of unitaries, such that this expression evaluates to a product of loop contractions. While this factor can in principle be obtained by keeping track of all contractions, it can also be obtained by observing that the final expression no longer depends on the choice of unitaries. As such, we can e.g. choose all gates to be equal to the identity to obtain this value. In this case the tensor network evaluates to
\begin{align}
    \figeq[0.25\columnwidth]{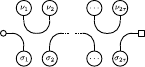} = \prod_{i=0}^{\tau} d^{|\sigma_{2i}^{-1} \sigma_{2i+1}|} \times \prod_{i=1}^\tau d^{|\nu_{2i-1}^{-1}\nu_{2i}|} =d \prod_{i=1}^{\tau} d^{|\sigma_{2i-1}| -|\sigma_{2i}|+k} \times \prod_{i=1}^\tau d^{|\nu_{2i}|-|\nu_{2i-1}|+k}.
\end{align}
where we have formally identified $\sigma_{0} = \circ$ and $\sigma_{2\tau+1} = \smallsquare$ and used $|\smallsquare|=1$. Plugging this expression in Eq.~\eqref{eq:app:identity_matel}, we only need to perform the summation over M\"obius functions to return the presented result. \\

To conclude, we briefly discuss one subtlety. The spatial transfer matrix $\mathcal{T}$ is necessarily contracting, i.e. all its eigenvalues lie either on or inside the unit circle~\cite{bertini_operator_i_2020}. 
Furthermore, it follows from the presented proof that $(\mathcal{I}|\mathcal{T}|\,\mathcal{I}) = (\mathcal{I}|\,\mathcal{I})$, since the influence matrix acts within the space of multichains and is an eigenstate when projecting $\mathcal{T}$ in this space. If $\mathcal{T}$ was Hermitian, the combination of these arguments would suffice to establish $|\mathcal{I})$ as an exact eigenstate of the transfer matrix for {\it any} choice of unitary gates: the equality $(\mathcal{I}|\mathcal{T}|\,\mathcal{I}) = (\mathcal{I}|\,\mathcal{I})$ could only be satisfied if $\mathcal{T}|\mathcal{I}) = |\mathcal{I})$. Any term in $\mathcal{T}|\,\mathcal{I})$ that is orthogonal to $|\mathcal{I})$ would increase the norm, contradicting the fact that $\mathcal{T}$ is contracting and hence its operator norm is upper bounded by 1. However, in the non-Hermitian case the matrix $\mathcal{T}$ can be contracting without having operator norm $1$, such that the previous argument does not apply. 
A numerical investigation indeed indicates that, away from the dual-unitary point, this transfer matrix has a unique leading eigenvalue $1$ and all other eigenvalues lie inside the unit circle, but it generically has singular values outside the unit circle.
We highlight this subtlety since otherwise our perturbative calculation would have established that $|\mathcal{I})$ is an exact eigenvector with unit eigenvalue for generic $\mathcal{T}$. While this is true at the dual-unitary point, it only holds approximately away from this limit.
\\

\section{Explicit results for the 2-OTOC}
\label{app:2OTOC}

Having obtained a proof of Eq.~\eqref{eq:app:identity} for general choices of $k$, we now illustrate this result for $k=2$. In this case there are two permutations, the identity $\circ$ and the cyclic permutation, i.e, the swap $\smallsquare$, and the multichains correspond to domain wall states, switching from $\circ$ to $\smallsquare$. These states have been extensively analyzed in the literature and can be explicitly orthonormalized to obtain the projector on this space~\cite{bertini_operator_i_2020,claeys_maximum_2020,rampp_dual_2023}. \\

{\bf Influence matrix.} For $k=2$, the multichains $|\vec{\sigma})$ are domain walls between the identity and swap permutation. In contrast to the main text, in this section, we use normalized domain wall states for convenience
\begin{align}
    |2\tau,j) = \frac{1}{d^{2\tau}}|\underbrace{\circ \dots \circ }_{j} \, \underbrace{\smallsquare \dots \smallsquare }_{2\tau-j}\,).
\end{align}
The influence matrix consists of a linear combination of these states.
For these permutations, the only nontrivial value the M\"obius function can take is given by $\mu(\circ,\smallsquare) = -1$.
The random matrix influence matrix is hence given by
\begin{equation}
   |\mathcal{I}) = \sum_{i=0}^\tau|2\tau,2i) -\frac{1}{d}\sum_{i=0}^{\tau-1} |2\tau,2i+1).
\end{equation}
We can first check that this state corresponds to the right boundary projected on the space of multichains, i.e. that the influence matrix satisfies $(2\tau,j|\,\mathcal{I}) = (2\tau,j|\pairs)$.
The overlap of these multichains with the right boundary state can be directly calculated as
\begin{equation}\label{eq:app:2OTOC:overlaps_R}
    (2\tau,j|\pairs) = \begin{cases}
        1,  &j\,\,\textrm{even},\\ 
        \frac{1}{d}, &j\,\,\textrm{odd}.
    \end{cases}
\end{equation}
Next we compute the overlaps of $|\mathcal{I})$ with individual multichains. As a warmup, we begin with the uniform state $|2\tau,0)$, for which
\begin{equation}
    (2\tau,0|\,\mathcal{I}) = \sum_{i=0}^\tau (2\tau,0|2\tau,2i) -\frac{1}{d}\sum_{i=0}^{\tau-1}(2\tau,0|2\tau,2i+1) = 1 + \sum_{i=1}^\tau\frac{1}{d^{2i}} - \frac{1}{d} \sum_{i=0}^{\tau-1}\frac{1}{d^{2i+1}} = 1.
\end{equation}
The latter two terms cancel exactly, giving the expected result. 
For general even $j$ we split the sums into parts $i<j$ and $i>j$,
\begin{equation}
    (2\tau,j|\,\mathcal{I}) = \sum_{i=0}^{j/2-1}\frac{1}{d^{j-2i}} + 1 + \sum_{i=j/2+1}^\tau\frac{1}{d^{2i-j}} - \frac{1}{d}\sum_{i=0}^{j/2-1}\frac{1}{d^{j-2i-1}} - \frac{1}{d}\sum_{i=j/2}^{\tau-1}\frac{1}{d^{2i+1-j}} = 1.
\end{equation}
The first sum cancels with the third and the second sum cancels with the fourth, such that the only remaining term returns the expected result $(2\tau,j|\,\mathcal{I}) = 1$. For odd $j$ we similarly find that
\begin{equation}
    (2\tau,j|\,\mathcal{I}) = \sum_{i=0}^{(j-1)/2}\frac{1}{d^{j-2i}} + \frac{1}{d} + \sum_{i=(j+1)/2+1}^\tau\frac{1}{d^{2i-j}} - \frac{1}{d}\sum_{i=0}^{(j-1)/2}\frac{1}{d^{j-2i-1}} - \frac{1}{d}\sum_{i=(j+1)/2}^{\tau-1}\frac{1}{d^{2i+1-j}} = \frac{1}{d}.
\end{equation}
The summations again cancel, such that only $(2\tau,j|\,\mathcal{I}) =1/d$ remains. These calculations reproduce the expected overlaps from Eq.~\eqref{eq:app:2OTOC:overlaps_R}, confirming that the random matrix influence matrix returns the correct influence matrix for the boundary scrambler when $k=2$. \\

{\bf Perturbative stability.}
Next, we verify that this influence matrix is an eigenstate with unit eigenvalue of the transfer matrix projected in the space of multichains.
For this we explicitly compute $(2\tau,j|\mathcal{T}|\,\mathcal{I})$. The calculation proceeds in a manner analogous to the calculation of $(2\tau,j|\,\mathcal{I})$, but because of the involved gates no longer being dual unitary, the individual terms cannot be reduced to mere powers of $1/d$ anymore. Instead, they are of the form
\begin{align}
    (2\tau,j|\mathcal{T}|2\tau,i) = \frac{1}{d^{4\tau+1}} \,\, \vcenter{\hbox{\includegraphics[width=0.385\columnwidth]{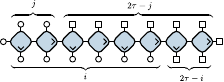}}}\,\,.
\end{align}
All such matrix elements can be expressed in terms of quantities
\begin{align}
    B_i = \frac{1}{d^{2i+1}}\,\, \vcenter{\hbox{\includegraphics[width=0.205\columnwidth]{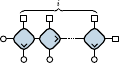}}}\,\,,
\end{align}
which depend on the details of the chosen gate and can generally not be further simplified. For convenience, we define $B_0=1$.
In order to arrive at these quantities, we note that unitarity can be used to simplify the expression for the matrix elements from either the left and the right. Unitarity results in the graphical identities
\begin{align}
    \vcenter{\hbox{\includegraphics[width=0.065\columnwidth]{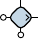}}}\,=\,\vcenter{\hbox{\includegraphics[width=0.065\columnwidth]{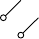}}}\,\,, \qquad 
    \vcenter{\hbox{\includegraphics[width=0.065\columnwidth]{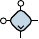}}}\,=\,\vcenter{\hbox{\includegraphics[width=0.065\columnwidth]{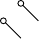}}}\,\,, \qquad
    \vcenter{\hbox{\includegraphics[width=0.065\columnwidth]{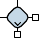}}}\,=\,\vcenter{\hbox{\includegraphics[width=0.065\columnwidth]{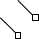}}}\,\,, \qquad
    \vcenter{\hbox{\includegraphics[width=0.065\columnwidth]{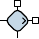}}}\,=\,\vcenter{\hbox{\includegraphics[width=0.065\columnwidth]{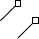}}}\,\,.
\end{align}
Using these identities, the matrix elements can be simplified as, e.g.,
\begin{align}
    \vcenter{\hbox{\includegraphics[width=0.385\columnwidth]{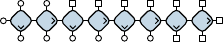}}}\,\,=\,\,\vcenter{\hbox{\includegraphics[width=0.315\columnwidth]{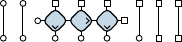}}}
\end{align}
For general choices of $i$ and $j$, these simplifications can be used to evaluate the matrix elements as
\begin{subequations}
\begin{align}
     &(2\tau,2j|\mathcal{T}|2\tau, 2i)  = \begin{cases}
         \frac{1}{d}B_{2j-2i-1}, & i<j,\\
         1, & i=j,\\
         \frac{1}{d}B_{2i-2j-1}, & i>j,
     \end{cases} &\quad
     &(2\tau,2j|\mathcal{T}|2\tau, 2i+1)  = \begin{cases}
         B_{2j-2i-1}, & i<j,\\
         B_{2i+1-2j}, & i>j,
     \end{cases} \\
     &(2\tau,2j+1|\mathcal{T}|2\tau, 2i)  = \begin{cases}
         \frac{1}{d}B_{2j-2i}, & i\leq j,\\
         \frac{1}{d}, & i=j+1,\\
         \frac{1}{d}B_{2i-2j}, & i>j+1,
     \end{cases} &\quad
     &(2\tau,2j+1|\mathcal{T}|2\tau, 2i+1)  = \begin{cases}
         B_{2j-2i}, & i<j,\\
         1, & i=j,\\
         B_{2i-2j}, & i>j.
     \end{cases} 
\end{align}
\end{subequations}
Using these matrix elements, we can now evaluate the action of the projected transfer matrix on the influence matrix.
While individual terms are more complicated, the previously observed cancellation between the even and odd domain wall states still persists, and can be used to establish Eq.~\eqref{eq:app:identity}. 
For domain walls on even bonds $j$, we have
\begin{equation}
    (2\tau,j|\mathcal{T}|\,\mathcal{I}) = \sum_{i=0}^{j/2-1}\frac{B_{j-2i-1}}{d} + 1 + \sum_{i=j/2+1}^\tau\frac{B_{2i-j-1}}{d} - \frac{1}{d}\sum_{i=0}^{j/2-1}B_{j-2i-1} - \frac{1}{d}\sum_{i=j/2}^{\tau-1}B_{2i+1-j} = 1,
\end{equation}
and on odd bonds
\begin{equation}
    (2\tau,j|\mathcal{T}|\,\mathcal{I})= \sum_{i=0}^{(j-1)/2}\frac{B_{j-2i-1}}{d} + \frac{1}{d} + \sum_{i=(j+1)/2+1}^\tau\frac{B_{2i-j-1}}{d} - \frac{1}{d}\sum_{i=0}^{(j-1)/2}B_{j-2i-1} - \frac{1}{d}\sum_{i=(j+1)/2}^{\tau-1}B_{2i+1-j} = \frac{1}{d}.
\end{equation}
Taken together, we find that $(2 \tau,j|\mathcal{T}|\,\mathcal{I}) = (2 \tau,j|\,\mathcal{I})$, returning the expected eigenvalue equation in the space of multichains.\\

{\bf Generator of the dynamical semigroup.}
In the following we present an alternative derivation of the dynamics of the 2-OTOC [Eq.~\eqref{eq:dynamical_2OTOC}] based on the Jordan decomposition of the matrix $\mathcal{G}$ by evaluating Eq.~\eqref{eq:2OTOC_semi_group} explicitly.
This derivation also helps to clarify how the resulting expression needs to be modified when considering the late-time dynamics of generic operators (rather than eigenoperators). 

We first note that, due to its upper triangular block structure, the spectrum of $\mathcal{G}$ follows from the spectra of the diagonal channels
$\mathcal{M}_{\circ\circ}$ and $\mathcal{M}_{\sq \sq}$.
Following the arguments in the main text, the spectra of those channels can in turn be obtained from the spectrum of the quantum channel $\mathcal{M}$ [Eq.~\eqref{eq:single_replica_channel}].
More precisely, denoting the eigenvalues of $\mathcal{M}$ by $\lambda_0=1,\lambda_1,\ldots, \lambda_{d^2-1}$, the eigenvalues of the channels $\mathcal{M}_{\circ\circ}$ and $\mathcal{M}_{\sq \sq}$ are the products $\lambda_i\lambda_j$.
Consequently, those are also the eigenvalues of $\mathcal{G}$, which have algebraic multiplicity two due to the two diagonal blocks.
For $i\neq j$ or $i=j=0$ these also have geometric multiplicity two, whereas otherwise they have geometric multiplicity one, leading to a
two-dimensional Jordan block for eigenvalues of the form $\lambda_i^2$. As argued in Ref.~\cite{fritzsch_free_2025}, these govern the late-time dynamics of generic 2-OTOCs. 

To obtain an explicit expression for the 2-OTOC we again focus on $a$ and $b$ being right and left eigenvectors of $\mathcal{M}$ for eigenvalue $\lambda$ with $\mathrm{tr}(ab)=d$.
We denote the corresponding left and right eigenvectors of $\mathcal{M}_{\circ\circ}$ with eigenvalue $\lambda^2$ by $(b_{\circ}|= (\medcirc|\left(b \otimes \mathbf{1}_0 \otimes b  \otimes \mathbf{1}_0  \right)$ and $|a_{\circ})=\left(a \otimes \mathbf{1}_0 \otimes a  \otimes \mathbf{1}_0  \right)|\medcirc)$, respectively, noting that $|a_{\circ})$ corresponds to $|\medblackcirc)$ in the main text.
The normalization is chosen such that $(b_{\circ}|a_\circ)=[\mathrm{tr}(ab)]^2 = d^2$.
Similarly, we denote the left and right eigenvectors of $\mathcal{M}_{\sq\sq}$ with eigenvalue $\lambda^2$ by $(b_{\sq}|=(\smallsquare|\left(b \otimes \mathbf{1}_0 \otimes b  \otimes \mathbf{1}_0  \right)$ and $|a_{\sq})=\left(a \otimes \mathbf{1}_0 \otimes a  \otimes \mathbf{1}_0  \right)|\smallsquare)$, respectively, such that $(b_{\sq}|$ corresponds to $(\smallblacksquare|$ in the main text. 
The normalization is such that $(b_{\sq}|a_{\sq})=\mathrm{tr}(ab)^2 = d^2$, whereas $(b_{\sq}|a_{\circ})=(a_{\sq}|b_{\circ})=\mathrm{tr}(abab)$.

We now consider the corresponding Jordan block of $\mathcal{G}$, which can be parametrized as
\begin{align}
    J_{\lambda^2} = \lambda^{2} \Big[|R_0)(L_1| + |R_1)(L_0)\Big] +  |R_0)(L_0|
\end{align}
with $|R_0)$ and $(L_0|$ right and left eigenvectors satisfying
\begin{align}
    \left(\mathcal{G}-\lambda^2\mathbf{1}\right)|R_0) = 0 \quad \textrm{and} \quad (L_0|\left(\mathcal{G}-\lambda^2\mathbf{1}\right) = 0,
\end{align} 
and $|R_1)$ and $(L_1|$ the corresponding generalized right and left eigenvectors satisfying
\begin{align}
\label{eq:app:left_right_gen_evecs}
    \left(\mathcal{G}-\lambda^2\mathbf{1}\right)|R_1) = |R_0) \quad \textrm{and} \quad (L_1|\left(\mathcal{G}-\lambda^2\mathbf{1}\right) = (L_0|.
\end{align}
The above states are subject to the biorthonormalization condition $(L_i|R_j)=1-\delta_{ij}$, i.e., left and right eigenvectors are orthogonal and so are their generalized counterparts. The biorthonormality between these (generalized) eigenvectors yields the powers of the Jordan block as
\begin{align}\label{eq:app:powersjordan}
    J_{\lambda^2}^t = \lambda^{2t} \Big[|R_0)(L_1| + |R_1)(L_0)\Big] + t\lambda^{2(t-1)} |R_0)(L_0|\,.
\end{align}

The eigenstates follow directly from the eigenoperators of the diagonal channels, i.e. $\mathcal{M}_{\sq\sq}|a_{\sq}) = \lambda^2 |a_{\sq})$ and $(b_{\circ}|\mathcal{M}_{\circ\circ} = \lambda^2 (b_{\circ}|$, yielding eigenstates of $\mathcal{G}$ as
\begin{align}
    |R_0)  = \left( \begin{matrix}
     \alpha{\sq}|a_{\sq}) \\
        0
    \end{matrix} \right)  
    \quad \text{and} \quad
    (L_0| = \left(0, \beta_\circ(b_{\circ}|/d^4  \right)
\end{align}
where the first and second entry corresponds to the $\smallsquare$ and $\medcirc$ block, respectively.
Here, $\alpha_{\sq}$ and $\beta_\circ$ are nonzero complex numbers to be determined. 
Writing out the generalized eigenvalue equation~\eqref{eq:app:left_right_gen_evecs} fixes the $\circ$ component of $|R_1)$ to be a right eigenvector of $\mathcal{M}_{\circ\circ}$ with eigenvalue $\lambda^2$, as given by $\alpha_{\circ} |a_{\circ})$ (again with an prefactor $\alpha_{\circ}$ to be determined).
Similarly, the $\smallsquare$ component of $(L_1|$ is forced to be a left eigenvector $\beta_{\sq} (b_{\sq}|$ of $\mathcal{M}_{\sq\sq}$ for some complex number $\beta_{\sq}$, and we write an ansatz for the generalized eigenvectors as 
\begin{align}
    |R_1)  = \left( \begin{matrix}
       | r_1 ) \\
        \alpha_{\circ} |a_{\circ})
    \end{matrix} \right)  
    \quad \text{and} \quad
    (L_1| = \left(\beta_{\sq} (b_{\sq}|, (l_1 | \right),
\end{align}
with a priori unknown vectors $|r_1)$ and $(l_1|$.
The generalized eigenvalue equation~\eqref{eq:app:left_right_gen_evecs} translates to conditions on $|r_1)$ and $(l_1|$, respectively, reading
\begin{align}
\label{eq:app:condition_generalized_right_evecs}
    \left(\mathcal{M}_{\sq\sq} - \lambda^2 \mathbf{1}\right)|r_1) + \alpha_{\circ} \left(\mathcal{M}_{\sq\circ} - \mathcal{M}_{\sq\sq}/d\right)|a_{\circ}) & = \alpha_{\sq}|a_{\sq}) \, , \\
     (l_1|\left(\mathcal{M}_{\circ\circ} - \lambda^2 \mathbf{1}\right) + \beta_{\sq} (b_{\sq} | \left(\mathcal{M}_{\sq\circ} - \mathcal{M}_{\sq\sq}/d\right) & =  \beta_{\circ}(b_{\circ}| / d^4 \, .
\end{align}
The above constitutes an underdetermined linear system for each $|r_1)$ and $(l_1|$ with either none or infinitely many solutions, depending on the choice of $\alpha_\circ$ and $\beta_{\sq}$. In order to obtain a solution, the latter can be fixed by projecting the above equations onto $(b_{\sq}|$ and $|a_\circ)$.
Using that $(b_{\sq}|\mathcal{M}_{\sq\sq} = \lambda^2(b_{\sq}|$ and $\mathcal{M}_{\circ\circ} |a_\circ) = \lambda^2 |a_\circ)$, we obtain
\begin{align}
\label{eq:app:fix_alpha_beta}
   \alpha_{\circ} (b_{\sq}|\left(\mathcal{M}_{\sq\circ} - \mathcal{M}_{\sq\sq}/d\right)|a_{\circ}) & = \alpha_{\sq}(b_{\sq}|a_{\sq}) = d^2 \alpha_{\sq} \, , \\
      \beta_{\sq} (b_{\sq} | \left(\mathcal{M}_{\sq\circ} - \mathcal{M}_{\sq\sq}/d\right)|a_\circ) & = \beta_{\circ}(b_{\circ}|a_\circ)/d^4 = \beta_{\circ}/d^2 \, .
\end{align}
The above fixes the ratios $\alpha_{\sq}/\alpha_{\circ}$ and $\beta_{\sq}/\beta_{\circ}$.
Further constraints arise from the biorthonormality conditions for the mixed terms, which read
\begin{align}
    1 = (L_1|R_0) =  \alpha_{\sq}\beta_{\sq}(b_{\sq}|a_{\sq}) = \alpha_{\sq}\beta_{\sq} d^2 \quad \text{and} \quad 1 = (L_0|R_1) = \alpha_{\circ} \beta_{\circ} (b_{\circ}|a_{\circ})/d^4 = \alpha_{\circ} \beta_{\circ}/d^2,
\end{align}
fixing $\alpha_{\circ}\beta_{\circ} = d^2$ and $\alpha_{\sq}\beta_{\sq}=1/d^{2}$.
Orthogonality is automatically fulfilled for the eigenvectors, since $(R_0|L_0)=0$, and for the generalized eigenvectors we find that
\begin{align}
   \label{eq:app:orthogonality_gen_evecs}
   0 = (L_1|R_1) = \beta_{\sq}(b_{\sq}|r_1) + \alpha_{\circ}(l_1|a_{\circ})\,.
\end{align}
Formally, $|r_1)$ and $(l_1|$ can be obtained through pseudo-inversion, fixing $(b_{\sq}|r_1)= (l_1|a_{\circ}) = 0$.
The relations above are now sufficient to compute the 2-OTOC between $a$ and $b$ from Eq.~\eqref{eq:app:powersjordan}. The overlaps of the (generalized) eigenstates with the boundary vectors,
\begin{align}
    (\psi_b| = \left( (b_{\sq}|/d^2\,, 0 \right) \quad \text{and} \quad
    |\psi_a) = \left( \begin{matrix}
       d\,\, |a_{\circ}) \\
       d^2 |a_{\circ}) 
    \end{matrix} \right) \, ,
\end{align}
can be straightforwardly calculated as
\begin{align}
    (\psi_b|R_0) & = \alpha_{\sq}(b_{\sq}|a_{\sq})/d^2 = \alpha_{\sq} \,,\qquad
    (\psi_b|R_1)  = (b_{\sq}|r_1)/d^2 = 0\, , \\
    (L_0|\psi_a) & =\beta_{\circ}(b_{\circ}|a_{\circ})/d^2 = \beta_{\circ} \, , \qquad
    (L_1|\psi_a)  = d\beta_{\sq}  (b_{\sq}|a_{\circ})  + d^2 (l_1|a_{\circ}) = d \beta_{\sq}\, \mathrm{tr}(abab) \, .
\end{align}
With these overlaps, we recover the result from the main text:
\begin{align}
\label{eq:app:2OTOC_Jordan_block}
    C_2(t) = (\psi_b|J_{\lambda^2}^{t} |\psi_a) = \lambda^{2t} \alpha_{\sq} \beta_{\sq} d\mathrm{tr}(abab)  + \alpha_{\sq}\beta_\circ t\lambda^{2(t-1)} = \lambda^{2t}\mathrm{tr}(abab)/d +  t\lambda^{2(t-1)} (b_{\sq} | \left(\mathcal{M}_{\sq\circ} - \mathcal{M}_{\sq\sq}/d\right)|a_\circ) \, .
\end{align}
For general initial operators only the overlaps need to be modified, indicating that the overall form of the 2-OTOC will still apply albeit with modified prefactors.\\

{\bf Frequency-resolved cumulants.} To conclude this section, we discuss the details for the frequency-resolved cumulants, where we follow the approach of Ref.~\cite{fritzsch_eigenstate_2021}.
For our choice of a Hermitian gate $U$ one has $C_k(t)=C_k(-t)$ and similar for the free cumulants.
Hence, we simply need to compute the discrete-time Fourier transform (DTFT) of $\lambda^t$ and $t \lambda^t$.
For convenience, we introduce the decay rate $\gamma$ and assume $\lambda$ positive (as in the main text), such that $\lambda^t \equiv e^{-\gamma t}$, which corresponds to $\gamma \equiv - \ln |\lambda| > 0$.
We thus have
\begin{equation}
    \label{eq:fourier_term_1}
    \sum_{t=-\infty}^\infty e^{-\gamma|t| + i\omega t}
    =
    1 + 
    \sum_{t=1}^\infty [
        e^{-(\gamma + i\omega)t}
        +
        e^{-(\gamma - i\omega)t}
    ]
    =
    \frac{1}{1 - e^{-\gamma -i\omega}}
    +
    \frac{1}{1 - e^{-\gamma + i\omega}}
    - 1
    =
    \frac{\sinh \gamma}{\cosh \gamma - \cos \omega}
    \equiv
    J^{(1)}_\gamma(\omega),
\end{equation}
which corresponds to the expression of the main text.
This result hence describes the two-point correlation function in the frequency domain.
We can similarly compute the DTFT of $t \lambda^t = t e^{-\gamma t}$ by taking the derivative of the result above, using that $- d_{\gamma } (e^{-\gamma |t|}) = |t| e^{-\gamma |t|}$, to find
\begin{equation}
    \label{eq:fourier_term_2}
    \sum_{t=-\infty}^\infty |t| e^{-\gamma|t| + i\omega t}
    =
    -
    \frac{d}{d\gamma}
    \left(
    \frac{\sinh \gamma}{\cosh \gamma - \cos \omega}
    \right)
    =
    \frac{\cos(\omega)\cosh(\gamma) - 1}{\left[\cos(\omega) - \cosh(\gamma)\right]^2}
    \equiv
    J^{(2)}_\gamma(\omega),
\end{equation}
for the second contribution, which has a large negative peak at $\omega = 0$ and two smaller neighboring peaks at intermediate frequencies.
For the terms $\lambda^{2t}$ and $t \lambda^{2t}$ appearing in the analytical expression for the 2-OTOC, we simply substitute $\gamma \rightarrow 2\gamma$ in Eqs.~\eqref{eq:fourier_term_1} and~\eqref{eq:fourier_term_2}.

\section{Details on the numerical implementation}
\label{app:numerics}

{\bf Choice of unitary.}
For the numerics of Fig.~\ref{fig:otoc_plot}, we choose the unitary $U$ as
\begin{align}
U = \begin{pmatrix}
0.7214 & 0.3618 - 0.0674\mathrm{i} & -0.4365 - 0.1884\mathrm{i} & 0.2964 + 0.1740\mathrm{i} \\
0.3618 + 0.0674\mathrm{i} & 0.5139 & 0.5213 + 0.3501\mathrm{i} & -0.3428 - 0.2977\mathrm{i} \\
-0.4365 + 0.1884\mathrm{i} & 0.5213 - 0.3501\mathrm{i} & 0.1888 & 0.5821 + 0.0723\mathrm{i} \\
0.2964 - 0.1740\mathrm{i} & -0.3428 + 0.2977\mathrm{i} & 0.5821 - 0.0723\mathrm{i} & 0.5759
\end{pmatrix}
\end{align}
This unitary, while randomly generated, has two properties that help to simplify the presented numerics. First, it is Hermitian, such that the left and right eigenoperators of the quantum channel $\mathcal{M}$ are identical (Hermitian, traceless) operators. For this reason, we can obtain exact results for higher-order correlations between matrix elements of a single operator. 
The corresponding advantage of having equal left and right eigenoperators is the fact that the $k$-OTOCs are symmetric in time, i.e. $C_k(t) = C_k(-t)$, which allows us to compare the frequency-resolved cumulants under Eq.~\eqref{eq:eth_cumulants} with the exact results.
Second, the quantum channel $\mathcal{M}$ acts as a projector on this eigenoperator: it only has a single nontrivial eigenvalue $\lambda \approx 0.9$, where fixing $\lambda$ close to $1$ results in slower OTOC dynamics and hence a clearer signal for longer times. \\

{\bf Dual-unitary qubit gates.}
Here we introduce the parametrization we use to implement the dual-unitary gates $V$ in the effective bath, as discussed in Sec.~\ref{sec:model}.
For convenience, we make use of the Cartan decomposition for two-qubit gates~\cite{khanejaCartanDecompositionSU2n2001, khanejaTimeOptimalControl2001, krausOptimalCreationEntanglement2001}, where we have:
\begin{equation}\label{eq:app:du_gate_parametrization}
    V 
    = 
    (v_+ \otimes v_-)
    e^{
    -i \tau
    \left( X \otimes X 
    +  Y \otimes Y 
    + J_z\, Z \otimes Z 
    \right)
    }
    (u_+ \otimes u_-).
\end{equation}
The operators $X, Y, Z$ are the usual Pauli matrices, $u_\pm, v_\pm \in SU(2)$ are local single-qubit gates, $J_z \in \mathbb{R}$ is the anisotropy parameter and $\tau$ is the Trotter step, returning a dual unitary gate at $\tau = \pi/4$~\cite{bertini_exact_2019}.
For $\tau = 0$ the gate is purely local and generates no entanglement.
In our numerics, $J_Z$ is chosen randomly in the interval $J_z \in [0, 1)$, the single-qubit gates are randomly generated, and the Trotter step is $\tau = \pi/4$ throughout the main text.
\\

{\bf Frequency-resolved ETH cumulants.}
For completeness, we also discuss some subtleties in the implementation and calculation of ETH cumulants in the frequency domain.
These are defined as
\begin{align}
    \label{eq:eth_cumulants_freq}
    k_2(\omega) 
     = 
    \frac{1}{D}\sum_{i\neq j}A_{ij}B_{ji}\, 
    \delta^{(2\pi)}(\omega - \omega_{ij}) 
    \, , \qquad
    k_4(\omega) 
     = 
    \frac{1}{D}\sum_{i\neq j \neq k \neq l}A_{ij}B_{jk}A_{kl}B_{ji}\,
    \delta^{(2\pi)}(\omega - \omega_{ij} - \omega_{kl}) 
    \,,
\end{align}
where $\delta^{(2\pi)}(\omega) \equiv \sum_{n=-\infty}^\infty \delta(\omega + 2\pi n)$ denotes the Dirac comb.
For numerical purposes these are implemented in terms of a smoothed Gaussian, where we numerically replace the Dirac deltas by smoothed Gaussians $\delta_\nu(\omega)$ with mean zero and variance $1/\nu$, which corresponds to a sharply peaked Gaussian at zero for sufficiently large $\nu$. 
In our numerics, it suffices to choose $\nu = 20$.
Note that in the definition of cumulants for multi-time correlation functions the frequency-resolved cumulants can also be defined as~\cite{pappalardi_full_2025}:
\begin{equation}
    \label{eq:alternative_k4}
    \tilde{k}_4(\omega_1,\omega_2)
    = 
    \frac{1}{D}\sum_{i\neq j \neq k \neq l}A_{ij}B_{jk}A_{kl}B_{ji}\,
    \delta^{(2\pi)}_\nu(\omega_1 - \omega_{ij})
    \delta^{(2\pi)}_\nu(\omega_2 - \omega_{kl}) 
    \,,
\end{equation}
which only returns the above free cumulant after a convolution. 

\begin{figure}
    \centering
    \includegraphics[width=0.8\textwidth]{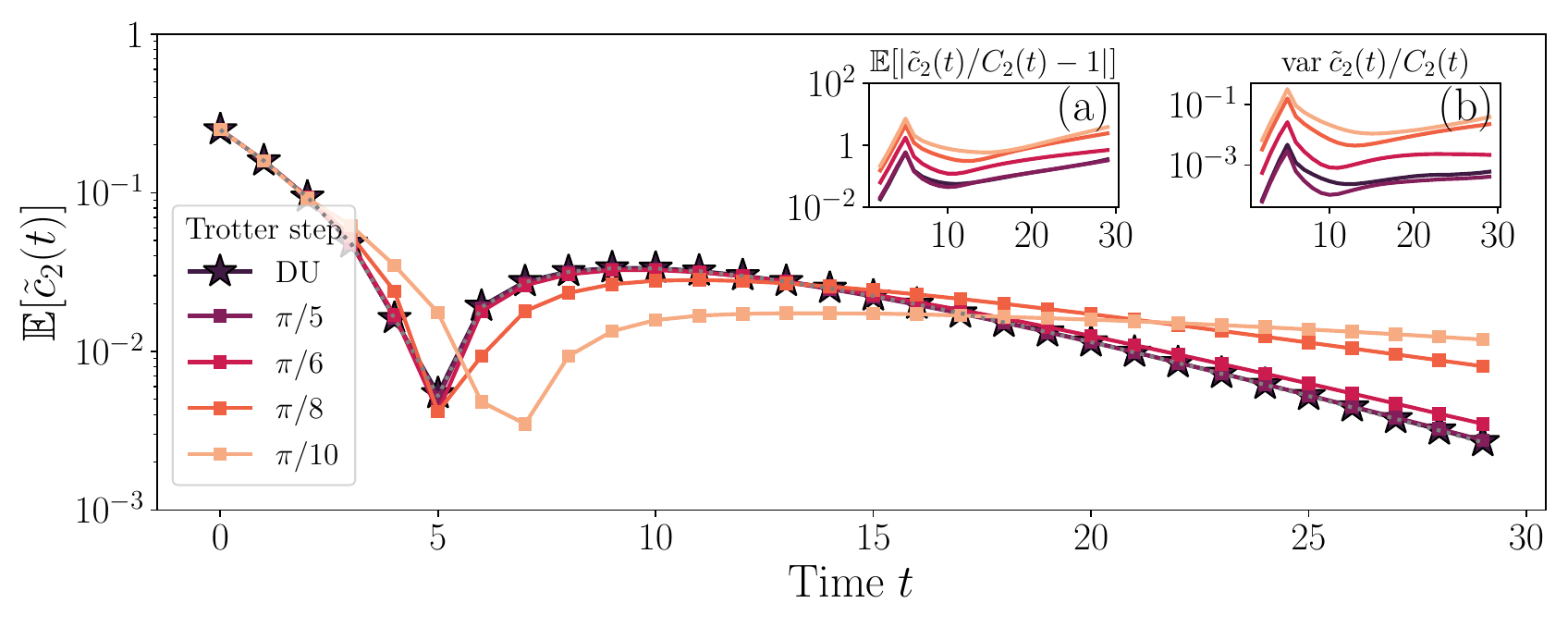}
    \caption{
    Comparison of numerical realizations $
    \tilde{c}_2(t)$ of the (absolute value of the) 2-OTOC with the analytical prediction $C_2(t)$ [Eq.~\eqref{eq:dynamical_2OTOC}] when moving away from dual-unitarity.
    Note the log-scale in the plot.
    We follow the parametrization from Eq.~\eqref{eq:app:du_gate_parametrization}, with decreasing Trotter step, ranging from $\tau = \pi/4$ (dual-unitary point, dark purple star markers) up to $\pi/10$ for generic unitary gates (square markers).
    In the insets (a) and (b), respectively, we also plot the average relative deviation and variance of the numerical realizations as a function of time.
    Simulations were performed for circuit size $L = 10$ and averaged over $400$ independent realizations.
    Further numerical details are described in App.~\ref{app:numerics}.
    }
    \label{fig:otoc_stability}
\end{figure}

\section{Numerical analysis of the perturbative stability}
\label{app:stability_numerics}

This section provides a numerical complement to App.~\ref{app:2OTOC} and the discussion on perturbative stability therein. 
In Fig.~\ref{fig:otoc_stability} we numerically investigate the perturbative stability of the influence matrix.
For that purpose, we parametrize the unitaries in the bath through the Trotter step $\tau$, per Eq.~\eqref{eq:app:du_gate_parametrization}.
At $\tau = \pi/4$ the gates are dual-unitary, with deviations from this value quantifying the deviation from dual-unitarity.
We denote the numerical and analytical values for the 2-OTOC as $\tilde{c}_2(t)$ and $C_2(t)$, respectively.
Once the Trotter step is chosen, the bath (zig-zag circuit $\mathcal{V}$) is constructed with random and independent unitaries with fixed $\tau$, as depicted by the blue unitaries in the bulk in Fig.~\ref{fig:intro}.
The (green) unitary $U$ at the boundary, which determines the quantum channel~\eqref{eq:single_replica_channel} and the 2-OTOC~\eqref{eq:dynamical_2OTOC} is also fixed and chosen to be the same as the one in the main text.
In order to remove the dependence on the specific choice of circuit, we additionally average over different realizations of the circuit.
The analytical prediction $C_2(t)$ for the dual-unitary point as shown in Fig.~\ref{fig:otoc_stability} is thus the same as the one in the main text and is depicted with the gray dotted line.
For the time interval we consider, the DU bath has excellent agreement with the analytical prediction, even at finite system size.
Tuning away from the dual-unitary point, the OTOC is visually indistinguishable from the exact result even at Trotter step $\tau = \pi/5$, which is non-dual-unitary. Only at $\tau = \pi/6$ do we start to see some small deviations, specially at longer times. The further away from the the dual-unitary points, the more the dynamics deviate from the analytical prediction, but importantly it does so in a smooth manner.

We also introduce two quantities as helpful metrics. First, we consider the relative deviation of a numerical finite-size realization, given by:
$
    \mathbb{E}[|
    {\tilde{c}_2(t)}/{C_2(t)} - 1
    |],
$
where $\mathbb{E}[ \bullet ]$ denotes the ensemble average over different realizations.
There are shown in Fig.~\ref{fig:otoc_stability}~[Inset (a)], and qualitatively they follow the behavior we described in the previous paragraph.
The small peak at $\tau \approx 5$ occur due to the sign-change of the 2-OTOC, where the correlations at that point are close to zero.
We can see that the relative deviation increases with time, as finite-size effects become starker.

Similarly, we also investigate fluctuations through the corresponding (relative) variance $\mathrm{var\: } \tilde{c}_2(t)/C_2(t)$ for the finite-size correlations.
Analogously to the correlations themselves, the fluctuations [Inset (b)] are minimal at or near the dual-unitary point ($\tau = \pi/4$ and $\tau = \pi/5$), and they become more apparent as we move further away from dual-unitarity.
Overall, the fluctuations follow the same qualitative behavior as the relative difference of the correlations seen in Fig.~\ref{fig:otoc_stability}~[Inset (a)].\\

\begin{figure}
    \centering
    \includegraphics[width=0.8\textwidth]{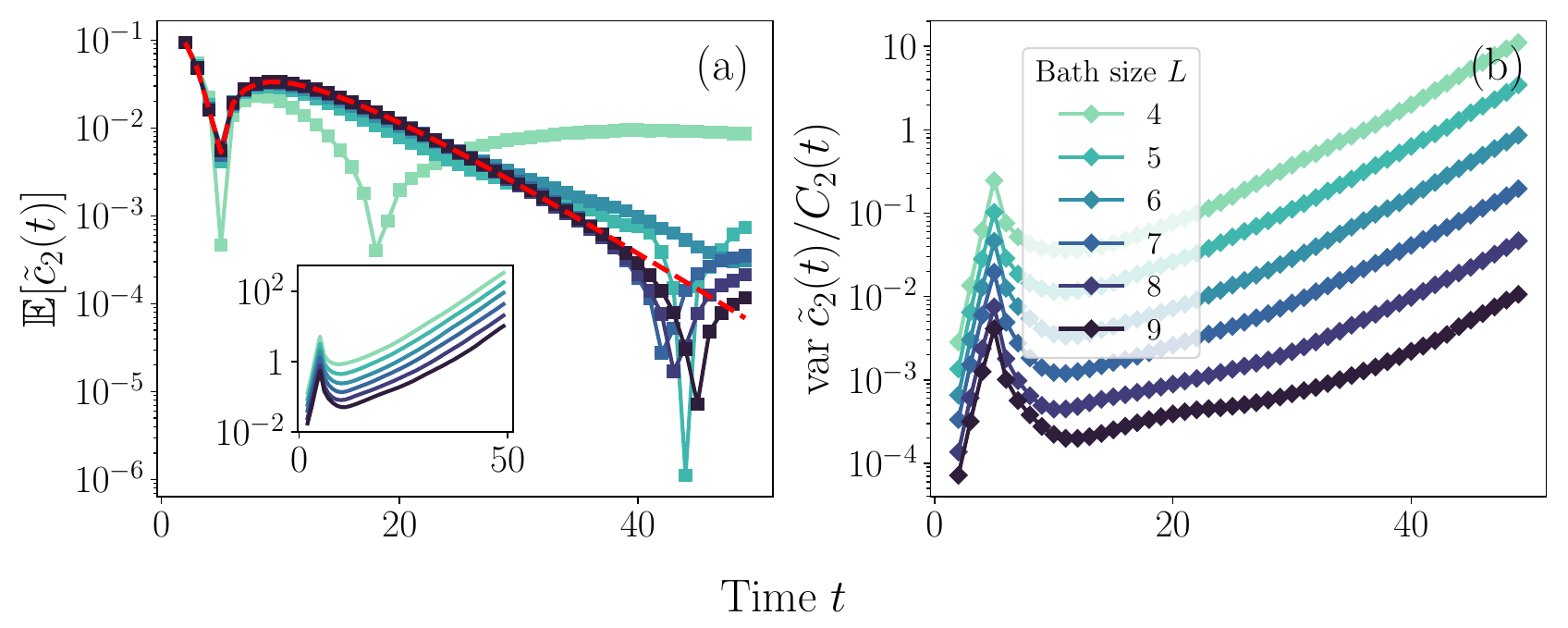}
    \caption{
    Numerical analysis for typicality of a finite-size realization in the boundary scrambler.
    (a) Numerical 2-OTOC (square markers) as function of time for different values of $L$. 
    Darker colors indicate larger sizes.
    We compare the numerics with the analytics plotted with the red dashed line for comparison.
    We plot the corresponding relative deviation on the inset.
    We can see that in general the accuracy exponentially improves with the bath size.
    Conversely, it decreases for larger times.
    (b) We analyze the fluctuations as a function of time and bath size.
    The behavior is analogous to the relative deviations, indicating concentration for larger bath dimension and shorter times.
    Simulations were performed over $2000, 1500, 1500, 1000, 1000, 500$ realizations for $L = 4, 5, 6, 7, 8, 9$.
    }
    \label{fig:otoc_concentration}
\end{figure}

\section{Approach to the thermodynamic limit}
\label{app:concentration}

In this section we perform numerical investigations on the approach to the thermodynamic limit and the dependence on the choice of dual-unitary gates, i.e. on the typicality of a finite-size realization.
The numerics explicitly shows an exponential convergence scaling as $1/D$ towards the thermodynamic limit, with fluctuations over different realizations suppressed as $1/D^2$, as we show in Fig.~\ref{fig:otoc_concentration}.
In Fig.~\ref{fig:otoc_concentration}~(a) we compare the numerics of the 2-OTOC $\tilde{c}_2(t)$ with the analytical prediction (red dashed line).
This is shown in more detail through the average relative deviation in the inset, as defined in App.~\ref{app:2OTOC}.
For bath size $L = 7$ we already observe very good agreement with the analytics up to $t \approx 35$.
Similarly, in panel~(b) we see a very similar behavior for the fluctuations.
By increasing the bath size, we observe both an exponential suppression of fluctuations and also exponentially smaller deviations for the analytical prediction at the thermodynamic limit.
In the same spirit, deviations and fluctuations become larger for longer times.
Nevertheless, fluctuations display a certain stability for intermediate times, and start to increase only after a certain threshold. 
This is specially clear for $L_E = 7$ in Fig.~\ref{fig:otoc_concentration}~(b), where the relative variance is constant up to $t \lesssim 25$.
Taken together, these results indicate that the analytical results approach the analytical result for the thermodynamic exponentially, with only a weak dependence on the choice of realization.

\end{document}